\documentclass[conference]{IEEEtran}
\IEEEoverridecommandlockouts
\usepackage{hyperref}
\usepackage{cite}
\usepackage{multirow}
\usepackage{amsmath,amssymb,amsfonts}
\usepackage{algorithmic}
\usepackage{graphicx}
\usepackage{textcomp}
\usepackage{mathtools}
\usepackage{multicol}
\usepackage{xcolor}
\usepackage{multicol}
\usepackage{comment}

\usepackage{booktabs}
\usepackage[accsupp]{axessibility}  
\usepackage{soul}
\usepackage{bm}
\usepackage{subcaption}

\usepackage{pgf}
\usepackage{tikz}
\usepackage{tikzscale}
\usetikzlibrary{arrows,external}
\usetikzlibrary{spy}
\usepackage{pgfplots}
\usepackage[skins]{tcolorbox}
\usepackage{import}
\usepackage{url}
\usepackage{tabu}

\usepackage{cite}  
\usepackage{booktabs}
\usepackage{pgf-pie}
\usepackage{pgfplots}
\usepackage{tikz}
\usepackage{amssymb}
\usepackage{xcolor}
\usepackage{soul}
\usepackage{hhline}

\usepackage{pgfplotstable}
\usepackage{caption}
\usepackage{adjustbox}

\pgfplotsset{compat=1.18}

\usepgfplotslibrary{groupplots}

\usepackage{multirow}
\usepackage{booktabs}
\usepackage{graphicx}
\usepackage{lscape}
\usepackage{graphicx}
\usepackage{array} 
\usepackage{pifont}
\usepackage{amssymb}
\usepackage{amsmath}
\usetikzlibrary{patterns}
\newcommand{\cmark}{\ding{51}} 

\usepackage{tikz}
\usetikzlibrary{shapes,arrows}

\def\BibTeX{{\rm B\kern-.05em{\sc i\kern-.025em b}\kern-.08em
    T\kern-.1667em\lower.7ex\hbox{E}\kern-.125emX}}
\begin{document}
\title{Sketch Animation: State-of-the-art Report}
\author{
\IEEEauthorblockN{Gaurav Rai, Ojaswa Sharma}
\IEEEauthorblockA{
Graphics Research Group, IIIT Delhi\\
gauravr@iiitd.ac.in, ojaswa@iiitd.ac.in
}
}
\maketitle
\begin{abstract}
Sketch animation has emerged as a transformative technology, bridging art and science to create dynamic visual narratives across various fields such as entertainment, education, healthcare, and virtual reality. This survey explores recent trends and innovations in sketch animation, focusing on methods that have advanced the state of the art. The paper categorizes and evaluates key methodologies, including keyframe interpolation, physics-based animation, data-driven, motion capture, and deep learning approaches. We examine the integration of artificial intelligence, real-time rendering, and cloud-based solutions, highlighting their impact on enhancing realism, scalability, and interactivity. Additionally, the survey delves into the challenges of computational complexity, scalability, and user-friendly interfaces, as well as emerging opportunities within metaverse applications and human-machine interaction. By synthesizing insights from a wide array of research, this survey aims to provide a comprehensive understanding of the current landscape and future directions of sketch animation, serving as a resource for both academics and industry professionals seeking to innovate in this dynamic field.
\end{abstract}

\section{Introduction}
\label{sec:first}

A sketch is a quick, freehand drawing used for the visual expression of ideas, objects, or scenes. Sketch animation plays a vital role in the creative process and technological innovation by enabling quick and expressive visualization of ideas and offering a way to create dynamic visual storytelling frameworks. It is widely used in movie production, education, and interactive design, allowing artists, designers, and animators to explore motion, timing, and scene composition without the complexity of detailed rendering. This makes it an essential tool for storyboarding, prototyping, and early concept development in fields like entertainment videos, game design, educational illustration, and user experience. Over the years, different sketch animation techniques have emerged; each having its advantages and challenges. \\
Traditional hand-drawn sketch illustrations are animated frame-by-frame~\cite{agarwala2004keyframe, terra2004performance, fu2011animated}. As computational techniques and interactive technologies advance, it evolves to support more varied and expressive forms.
Sketch animation has applications in various domains, including entertainment, education, storytelling, video editing, and virtual reality, that benefit conveying complex ideas in an accessible and engaging manner. 
Animation frameworks with procedural animation techniques~\cite{guay2015space, wu2023two} have seen significant advances over time. Many such approaches animate sketches using skinning methods with various control handles such as points, skeletons, and cages.
To further automate the process of animation, motion capture (MoCap) technologies~\cite{kitagawa2020mocap, kim2008drivenshape, schreiner2024adapt} have been explored to generate sketch animations by mapping captured motion data from body sensors to sketches. Physics-based simulation methods~\cite{agrawal2013diverse, frezzato2022synthesizing} produce realistic and natural motion effects in sketch animation. 
Furthermore, sketch interpolation methods~\cite{yang2017context, barroso2025automatic, shen2024bridging, jiang2022stroke, siyao2023deep, narita2019optical, zhu2025thin} reduce manual effort and overcome data dependency. 
In recent years, advancements in deep learning and generative AI have significantly transformed the field of sketch animation, making animation generation more accessible to novice users. 
Furthermore, generative AI techniques~\cite{sun2025spiritus, Zhong:2025:Sketch2Anim} animate sketches using generative models such as diffusion models~\cite{ho2020denoising, song2020denoising, blattmann2023stable}, guided by either video-based motion or text-based motion descriptions. These approaches allow users to customize motion according to their requirements, thereby expanding creative possibilities in sketch animation while enhancing the realism, fluidity, and responsiveness of the resulting animated sequences.

The use of Generative AI techniques amplifies the capabilities of sketch animation. It overcomes manual and traditionally labor-intensive processes, facilitating collaborative, scalable, and resource-efficient workflows. Despite these advancements, there are significant challenges in the field of sketch animation, such as computation complexity, trade-offs between quality, user control, accessibility, multi-user systems, and real-time interactive applications. Additionally, user-friendly sketch animation interfaces overcome the complexity for both novices and experts and provide freedom to access sketch animation technologies. Various sketch animation techniques, such as keyframe-based animation, autocomplete/interpolation, data-driven methods like motion capture, video motion retargeting, and text-driven animation, have continuously evolved with technological advancements to create more immersive and natural sketch animations.

This survey seeks to comprehensively map the evolving landscape of sketch animation by categorizing and critically evaluating key methodologies, highlighting technological innovations, and discussing the inherent challenges and future opportunities. This survey aims to serve as a valuable resource for academics, practitioners, and industry professionals.

\begin{figure*}[tbp]
    \centering
    \includegraphics[width=0.8\linewidth]{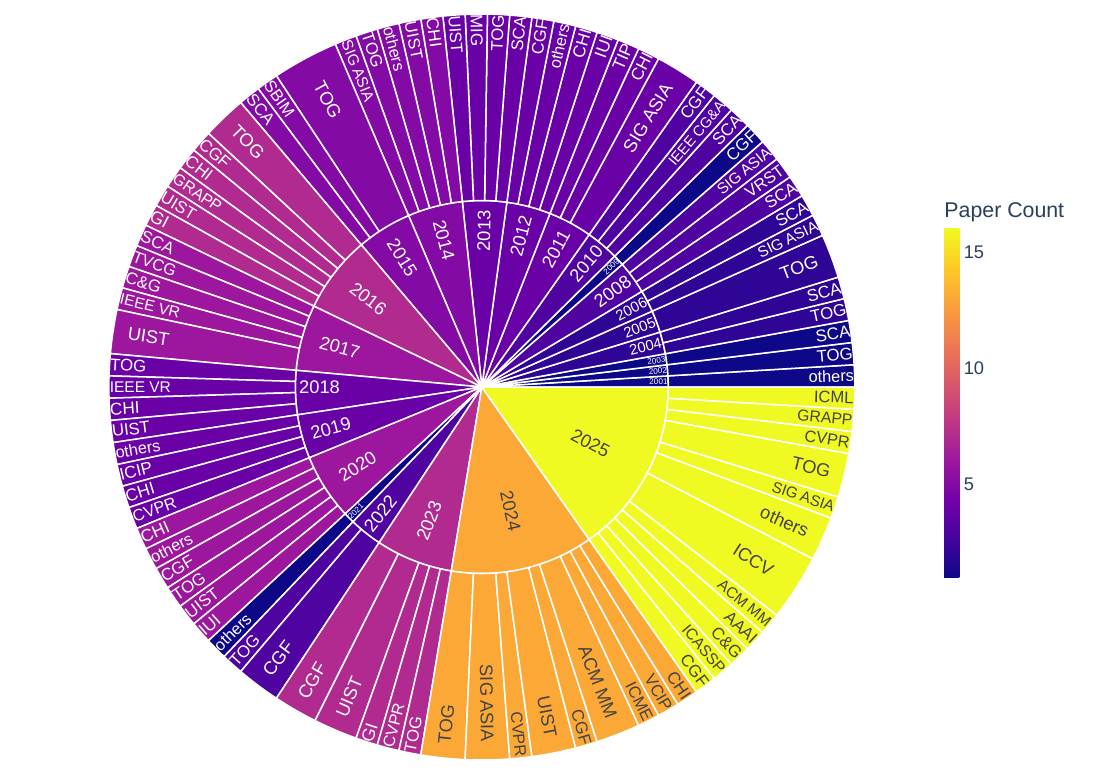} 
    \caption{Trends in sketch animation research from 2001 to 2025, showing the yearly number of published papers and the distribution of publications across venues over time.}
    \label{fig:yearly_dist}
\end{figure*}

\subsection{Evolution of sketch animation literature}
This survey presents a structured review of sketch animation methods, including their principles and applications. Our research found multiple surveys~\cite{liu2025state, schwartz2025computational, xu2022deep} conducted on sketches that cover various aspects of sketch processing and representation techniques for sketch animation. However, these works primarily address sketch analysis, understanding, and representation techniques, and do not comprehensively cover sketch animation. This limitation motivates our systematic review, which aims to bridge this gap by presenting a focused and structured survey of sketch animation research.

We have identified 105 relevant publications that contribute to the field of sketch animation during 25 years from 2001 until 2025. Fig.~\ref{fig:yearly_dist} present the distribution of these papers over the years and across various publication venues. 

\begin{figure*}[tbp]
    \centering
    \includegraphics[width=\linewidth]{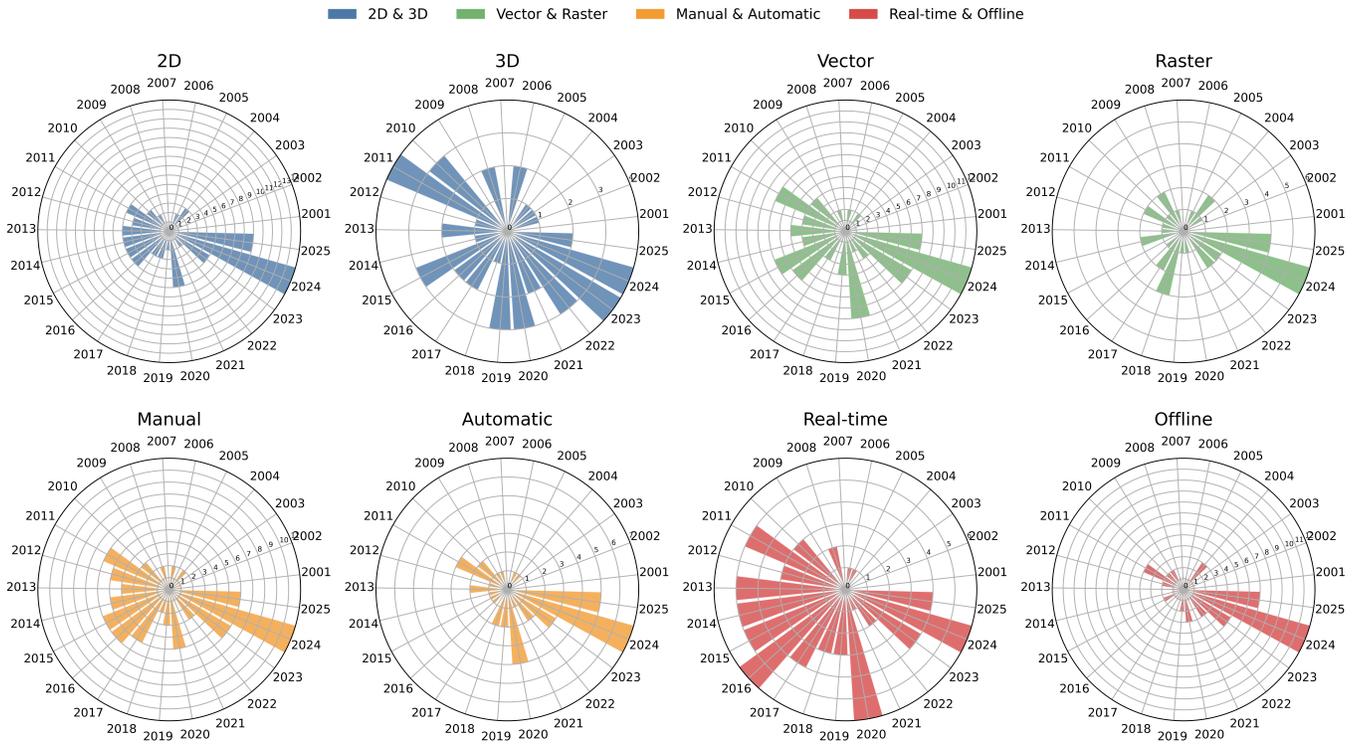} 
    \caption{Number of papers published from 2001–2025 on 2D/3D, raster/vector, automatic/manual, and real-time/offline.}
    \label{fig:charts}
\end{figure*}

We systematically categorized sketch animation methods based on several defining properties influencing their creation processes. One key distinction lies in the dimensionality of the animation, which can be either two-dimensional (2D) or three-dimensional (3D). 3D animation refers to animating 2D sketches within a 3D space, where the sketches are inflated or mapped onto volumetric forms or other 3D representations~\cite{feng2017magictoon, dvorovzvnak2020monster, weng2019photo, zhou2024drawingspinup}. Another axis of classification is sketch type, which includes vector-based sketches, characterized by resolution-independent (strokes) lines and shapes, and raster-based sketches, which rely on raster-based representations and are generally more suitable for capturing rich details and pixel-level control. Further, the classification of animations is based on the level of automation. We divided it into two categories: manual methods, where the user needs to provide additional input or suggestions to generate the animation, and automatic methods, which are fully automated and require no additional input or manual efforts. Sketch animations also differ in their computational process, such as real-time and offline. In real-time systems, the animation is generated dynamically in sync with a motion model (e.g., motion capture or video input). In contrast, offline methods rely on pre-defined motion rules or learned models (such as text-based inputs) to generate the animation offline. The evolution of sketch animation publications over time for these categories is illustrated in Fig.~\ref{fig:charts}.

\subsection{Contribution}
We provide a detailed overview of different sketch animation methodologies and their advancements over time. The method analysis given in Table~\ref{tab:comp_analysis} provides an insight into different techniques, input requirements, types of animation, and motion models. Further, Table~\ref{tab:analysis} describes the comparative analysis of different sketch animation techniques based on the animation properties such as automation, temporal consistency, flexibility, speed, expressiveness, etc. In particular in this review, we (a) provide a background of sketch animation by defining the importance of sketches and the significance of different sketch representations, (b) analyze different sketch animation methods and their recent advancements, (c) provide details on evaluation metrics for different types of sketch animation to measure animation quality, and (d) discuss applications and limitations of current sketch animation techniques along with future directions in this field.

\subsection{Structure of the review}
Our survey of sketch animation is structured as follows:
\begin{itemize}
    \item Section~\ref{sec:first} presents the introduction and overview, outlining the yearly publication trends and the key contributions of this survey.
    \item In section~\ref{sec:history}, we study the history of animation and the development of animation techniques and section~\ref{sec:background} describes the background, in which we discuss the different types of sketch representations.  
    \item Sections~\ref{sec:fourth}, \ref{sec:procedural}, \ref{sec:datadriven}, \ref{sec:2d3d}, and \ref{sec:interfaces} discuss methods for sketch animation, including inbetweening approaches for sketch animation, physics-based animation, data-driven (MoCap, video, text) methods, 2D sketch-to-3D animation, and sketch-animation interfaces. We describe the techniques used to animate sketches and outline their evolution.
    \item Section~\ref{sec:fifth} presents a comparative analysis of various sketch animation techniques focusing on their animation properties.
    \item Section~\ref{sec:sixth} explains various datasets and evaluation matrices used to measure sketch animation quantitatively.
    \item Section~\ref{sec:trends} describes recent trends in sketch animation methodology, focusing on generative AI, especially generative diffusion-based methods.
    \item Section~\ref{sec:application}, describes various applications of sketch animation for different tasks and analyzes applicability of various animation methods.
    \item Section~\ref{sec:future} combines the gaps, limitations, and future direction. In these subsections, we describe current gaps in various sketch animation methods and highlight their limitations. The survey comes to its conclusion in section~\ref{sec:conclusion}.
\end{itemize}

\section{A brief history of animation techniques}
\label{sec:history}
The history of animation outlines its roots from early optical experiments to the sophisticated digital techniques of the present. Traditional animation was created using the illusion of movement through hand-painted or sequential images with optical tools such as zoetropes or magic lanterns before the invention of modern cameras. 
These devices relied on persistence of vision, a perceptual phenomenon that allows the human eye to blend discrete images into a single moving scene. Initially, artists and inventors experimented with hand-painted sequences on glass slides or rotating drums, laying the foundation for later developments in animated storytelling. Animation evolved into a structured production process with the advent of film and photography.
Subsequently, cel animation introduced the practice of layering transparent sheets (cels) containing drawings over static backgrounds. The workflow evolved to include consistent registration across frames, along with tracing and inking on transparent cels placed over pencil roughs or painted backgrounds. Further, the development of keyframe animation techniques marked a significant milestone: animators sketch the primary poses through keyframes, then fill in the inbetweens by checking the motion before the final painting. This hierarchical process introduces the principles of timing and spacing, allowing animators to control the consistency and expressiveness of movement. Later, the technique enabled more systematic quality control, as animators could preview motion arcs using line tests before committing to final paint and photography. These techniques present fundamental principles in animation, including anticipation, squash and stretch, and follow-through, which remain essential in both traditional and digital animation.

Later, xerography reduced labor intensity by transferring line art directly onto cels, while the rotoscope projector enabled frame-by-frame tracing of live-action footage. Rotoscoping bridged the gap between live-action cinematography and drawn animation, controlling realistic and stylized works. As computing and digital interfaces evolved, animation moved into a new phase characterized by real-time interaction and intuitive control. Moscovich and Hughes~\cite{moscovich2001animation} introduce motion-by-example systems, where animators can manipulate objects or draw gestures directly on a digital interface, and the system records these motions for playback. This offers intuitive sketch animation for non-experts, though it currently lacks precision control compared to keyframe methods. Recent technological advancements have made more efficient and automatic solutions for sketch and extended sketch animation by preserving stroke topology, ensuring temporal consistency, and automating in-between frame generation while retaining the expressive qualities of hand-drawn strokes. 

Sketch animation has evolved from traditional keyframe and interpolation-based methods to advanced data-driven and generative approaches. Recent methods span motion capture, physics-based simulation, and deep learning, enabling more intuitive, flexible, and intelligent animation pipelines. In particular, the integration of generative AI leveraging diffusion models and multimodal prompts (e.g., video or text) has empowered novice users to produce smooth, expressive animations with minimal input. This survey examines recent advancements, highlighting how modern sketch animation methods enable controllable, realistic, and semantically rich animations that bridge the gap between user intent and dynamic motion synthesis.

\section{Background}
\label{sec:background}
Sketches are simplified and abstract representations of objects, characters, or scenes, depicted by line drawing, contours, raster sketch, and strokes that effectively convey essential structure and story~\cite{liu2025state, schwartz2025computational}. Sketches are inherently challenging to process due to their abstract and incomplete visual characteristics, such as the lack of texture, ambiguous or overlapping strokes, and discontinuous or broken contour lines, which make it difficult to reliably infer object boundaries and structural details. Unlike detailed illustrations, sketches focus on capturing proportions, movement, and the intent of the drawing over exact realism. It allows the flexibility to the artists or users to experiment with ideas, modify poses or perspectives, and convey actions or a story efficiently. Their flexible nature makes them highly suitable for creative pipelines that prioritize speed and adaptability.

\subsection{Sketch representations} 
\label{sec:sketch_representation}
Sketch representation is critical for interpreting and generating visual content or animation with minimal inference. Different sketch representations capture different aspects of visual information, offering various advantages and posing distinct challenges. The key representations of the sketches are vector and raster.

\begin{figure}[ht]
    \centering
    \includegraphics[width=\linewidth]{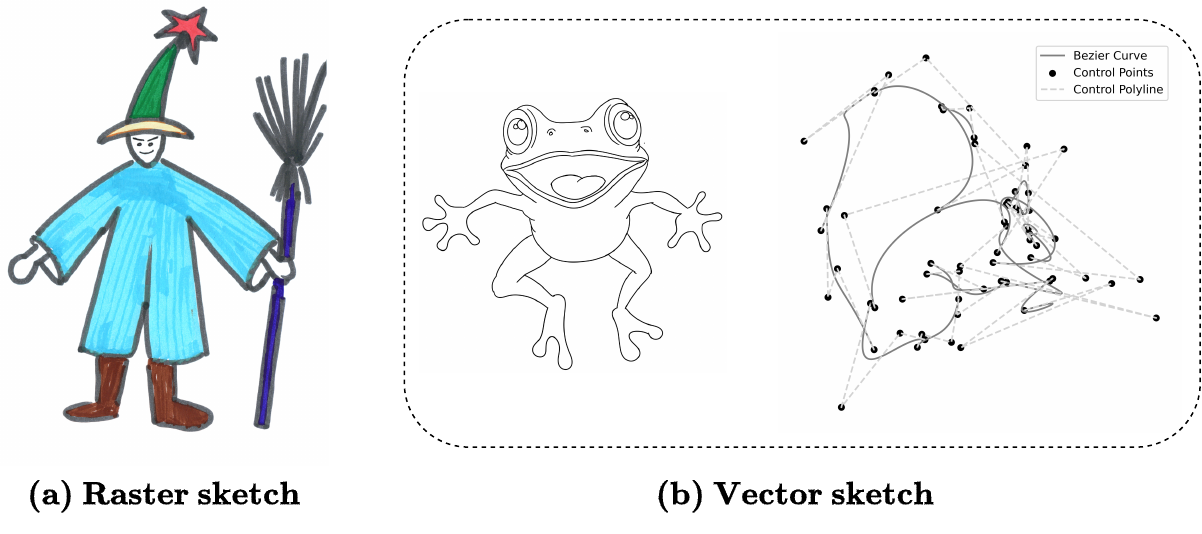}
    \caption{Raster and vector representations of the given sketches. Images adopted from~\cite{rai2024sketchanim, rai2024enhancing}.}
    \label{fig:sketch_rep}
\end{figure}

\paragraph{Vector sketches}
Vector sketches represent drawings as parameterized stroke sequences, such as polylines~\cite{li2020sketch}, B\'ezier curves~\cite{gal2024breathing} (see Fig.~\ref{fig:sketch_rep}(b)), or splines~\cite{banks1990real, kang2005interactive}. Unlike raster  sketches, vector sketches store drawing primitives in a resolution-independent format. The advantage of resolution independence is that it allows easy scaling and transformation without losing sketch quality. It is more suitable for modeling and editing fine-grained stroke properties and for neural network-based generation. The major challenge with vector sketches is that they are complex and cannot be generated accurately, especially from raster sketches. Handling artistic or highly abstract sketches where stroke order and structure vary widely is difficult. Vector sketches represent objects and scenes through continuous or discrete strokes. It is a simple, intuitive, and compact representation, suitable for many applications like retrieval, recognition, and generation. It is effective at capturing essential structural and shape information. 
Vector drawings can be ambiguous due to lack of texture and fine details, making it challenging to distinguish between visually similar classes based on their outlines.
In similar lines, temporal sketches make an add-on in vector sketches by incorporating the temporal sequence of strokes, which is how a drawing evolves. Each stroke has associated timestamps or sequential order, containing the artist's process. The advantage of temporal sketches is that they capture rich, dynamic information about the drawing process. It enables models to learn what was drawn and how it was drawn. The drawback is that temporal information can be noisy and requires large-scale, temporally annotated datasets.

\paragraph{Raster sketches}
Raster sketches can be multi-resolution, pixel-level abstractions or representations of visual concepts. They can contain texture information bounded by the silhouette, reducing complex shapes or layouts into a simplified grid-based structure as shown in Fig.~\ref{fig:sketch_rep}(a). Raster sketches are often labeled based on semantic classes. Their pixel-aligned format makes them suitable for convolutional neural networks~\cite{li_sketch-r2cnn_2018, smith2023method, rai2024sketchanim}. It demonstrates the significant performance improvements across several tasks, including sketch classification, sketch-based image retrieval, and sketch animation. The disadvantage is that the coarse resolution leads to inaccurate or overly simplistic object shapes and placements. Mapping continuous shapes onto a discrete grid often introduces distortions or unwanted artifacts.

In summary, while vector sketches offer a simple and direct abstraction to enhance flexibility and editing potential, temporal sketches enrich the representation with dynamic context and pixel-level abstractions or representations of visual concepts. However, each representation brings advantages and challenges and requires specialized processing techniques for robust sketch understanding and generation.

\begin{table*}[htbp]
\centering
\scalebox{1.0}{
\begin{tabular}{|l|ccc|cc|cc|cc|cc|l|}
\hline
\textbf{Sketch animation methods} & \multicolumn{3}{|c|}{\textbf{Additional inputs}} & \multicolumn{2}{c|}{\textbf{Input Types}} & \multicolumn{2}{c|}{\textbf{Dimension}} & \multicolumn{2}{c|}{\textbf{Interactivity}} & \multicolumn{2}{c|}{\textbf{Speed}} & \textbf{Motion model} \\
& \rotatebox{90}{Skeleton} & \rotatebox{90}{Video} & \rotatebox{90}{Text} & \rotatebox{90}{Vector} & \rotatebox{90}{Raster} & \rotatebox{90}{2D} & \rotatebox{90}{3D} & \rotatebox{90}{Manual} & \rotatebox{90}{Automatic} & \rotatebox{90}{Real-time} & \rotatebox{90}{Offline} &\\
\hline

\multicolumn{13}{|l|}{\textbf{Inbetweening}} \\
\hline
Davis et al.~\cite{davis2006sketching} & \cmark & & & \cmark & & \cmark &  & \cmark & & \cmark & & Pose / skeleton editing\\
BetweenIT~\cite{whited2010betweenit} & & & & \cmark & & \cmark & & \cmark & & & \cmark & Keyframe interpolation\\
Autocomplete~\cite{xing2015autocomplete} & & & & \cmark & & \cmark & & & \cmark & \cmark & & Stroke-based prediction\\
Corda et al.~\cite{corda2020real} & \cmark & & & \cmark & & \cmark & & & \cmark & \cmark & & Skeleton + cage deformation\\
Jiang et al.~\cite{jiang2022stroke} & & & & \cmark & & \cmark & & & \cmark & & \cmark & Data-driven inbetweening\\
Siyao et al.~\cite{siyao2023deep} & & \cmark & & & \cmark & \cmark & & & \cmark & & \cmark & Line inbetweening (raster)\\
Chen et al.~\cite{chen2023efficient} & & & & \cmark & & \cmark & & & \cmark & & \cmark & Frame interpolation\\
Brodt and Bessmeltsev~\cite{brodt2024skeleton} & \cmark & & & & \cmark & \cmark & & & \cmark & & \cmark & Bitmap + skeleton-aware\\
Mo et al.~\cite{mo2024joint} & & & & \cmark & & \cmark & & & \cmark & & \cmark & Joint stroke-based\\
Zhu et al.~\cite{zhu2025thin} & & & & \cmark & & \cmark & & & \cmark & & \cmark & Thin-plate / inbetweening\\
\hline

\multicolumn{13}{|l|}{\textbf{Physics-based methods}} \\
\hline
Zhu et al.~\cite{zhu2011sketch} & & & & \cmark & & \cmark & & & \cmark & \cmark & & Fluid simulation\\
Physink~\cite{scott2013physink} & & & & \cmark & & \cmark & & & \cmark & \cmark & & Rigid-body physics\\
Guay et al.~\cite{guay2015adding} & \cmark & & & \cmark & & & \cmark & & \cmark & \cmark & & Elastic line dynamics\\
Lingens et al.~\cite{lingens2020towards} & & & & & \cmark & \cmark & & & \cmark & & \cmark & Physics-based optimization\\

\hline
\multicolumn{13}{|l|}{\textbf{MoCap based techniques}} \\
\hline
MotionMaster~\cite{li2006motionmaster} & \cmark & & & & \cmark & \cmark & & \cmark & & & \cmark & MoCap-based transfer\\
TraceMove~\cite{patel2016tracemove} & \cmark & \cmark & & \cmark & & \cmark & & \cmark & & & \cmark & MoCap-based transfer\\
Pose2Pose~\cite{willett_pose2pose_2020} & \cmark & \cmark & & & \cmark & \cmark & & & \cmark & & \cmark & MoCap-based transfer\\
Animated Drawings~\cite{smith2023method} & & \cmark & & & \cmark & \cmark & & & \cmark & & \cmark & Data-driven mapping\\

\hline
\multicolumn{13}{|l|}{\textbf{Video-based motion transfer}} \\
\hline
Live Sketch~\cite{su2018live} & & \cmark & & \cmark & & \cmark & & \cmark & & & \cmark & Data-driven mapping\\
Wakey-wakey~\cite{xie2023wakey} & & \cmark & \cmark & & \cmark & \cmark & & & \cmark & & \cmark & Keypoint / conditional\\
Xie et al.~\cite{xie2024video} & & \cmark & & & \cmark & \cmark & & & \cmark & & \cmark & Video-based transfer\\
SketchAnim~\cite{rai2024sketchanim} & \cmark & \cmark & & & \cmark & \cmark & & & \cmark & \cmark & & Video-based transfer\\
SketchAnimator~\cite{deng2025animatesketches} & & \cmark & & \cmark & & \cmark & & & \cmark & & \cmark & T2V diffusion\\

\hline
\multicolumn{13}{|l|}{\textbf{Text-driven animation}} \\
\hline
BreathingSketches~\cite{gal2024breathing} & & & \cmark & \cmark & & \cmark & & & \cmark & & \cmark & T2V diffusion\\
DynamicTypography~\cite{liu2025dynamic} & & & \cmark & \cmark & & \cmark & & & \cmark & & \cmark & T2V diffusion\\
Rai and Sharma~\cite{rai2024enhancing} & & & \cmark & \cmark & & \cmark & & & \cmark & & \cmark & T2V diffusion\\
Flipsketch~\cite{bandyopadhyay2025flipsketch} & & & \cmark & \cmark & & \cmark & & & \cmark & & \cmark & T2V diffusion\\
AnimateSketches~\cite{deng2025animatesketches} & & & \cmark & \cmark & & \cmark & & & \cmark & & \cmark & T2V diffusion\\
FlexClip~\cite{khandelwal2025flexiclip} & & & \cmark & \cmark & & \cmark & & & \cmark & & \cmark & T2V diffusion\\
MoSketch~\cite{liu2025multi} & & & \cmark & \cmark & & \cmark & & & \cmark & & \cmark & T2V diffusion\\
GroupSketch~\cite{liang2025multi} & & & \cmark & \cmark & & \cmark & & & \cmark & & \cmark & T2V diffusion\\

\hline
\multicolumn{13}{|l|}{\textbf{2D sketch to 3D animation}} \\
\hline
MagicToon~\cite{feng2017magictoon} & \cmark & & & & \cmark & & \cmark & \cmark & & & \cmark & Skeleton-driven rigging\\
Photo-Wakeup~\cite{weng2019photo} & & \cmark & & & \cmark & & \cmark & & \cmark & & \cmark & Pose-driven rigging\\
MonsterMash~\cite{dvorovzvnak2020monster} & & & & \cmark & & & \cmark & & \cmark & \cmark & & Point-based deformation\\
DrawingSpinUp~\cite{zhou2024drawingspinup} & \cmark & & & & \cmark & & \cmark & & \cmark & & \cmark & MoCap-based transfer\\
Yoon et al.~\cite{yoon2025occlusion} & \cmark & & & & \cmark & & \cmark & & \cmark & & \cmark & MoCap-based transfer\\
Zhou et al.~\cite{zhou2025rigging} & \cmark & & & & \cmark & & \cmark & & \cmark & & \cmark & MoCap-based transfer\\
Smith et al.~\cite{smith2025animating} & \cmark & & & & \cmark & & \cmark & & \cmark & & \cmark & MoCap-based transfer\\

\hline
\multicolumn{13}{|l|}{\textbf{Sketch Animation tools / Interfaces}} \\
\hline
Kitty~\cite{kazi2014kitty} & & & & \cmark & & \cmark & & \cmark & & & \cmark & Interactive / rule-based\\
Draco~\cite{kazi2014draco} & & & & \cmark & & \cmark & & \cmark & & & \cmark & Interactive / heuristics\\
EnergyBrushes~\cite{xing2016energy} & & & &  & \cmark & \cmark & & \cmark & & \cmark & & Interactive / physics-guided\\
MotionAmplifier~\cite{kazi2016motion} & & & & \cmark &  & \cmark & & \cmark & & \cmark & & Interactive / rule-based\\
Mixed-Initiative~\cite{willett2018mixed} & & & & & \cmark & \cmark & & \cmark & & \cmark & & Mixed (user+auto)\\
RealityCanvas~\cite{xia2023realitycanvas} & & & & \cmark &  & \cmark & & \cmark & & \cmark & & Mixed (user + auto AR blending)\\
DrawTalking~\cite{rosenberg2024drawtalking} & & & & \cmark & & \cmark & & & \cmark & & \cmark & T2V / speech-guided\\
Augmented Physics~\cite{gunturu2024augmented} & & & & & \cmark & \cmark & & & \cmark & & \cmark & Physics-augmented synthesis\\
\hline
\end{tabular}
}
\caption{Comparison of prominent sketch-based animation methods.}
\label{tab:comp_analysis}
\end{table*}


\section{Inbetweening approaches for sketch animation}
\label{sec:fourth}
2D sketch animation concentrates on transforming static illustrations into dynamic sequences, allowing expressive freedom and controllable animation creation. Recent methods tackled the problem using geometry-based deformation techniques. 
The evolution of keyframe-based and inbetweening methods for animation aims to reduce the manual workload of generating intermediate frames while preserving artistic control, style coherence, and structural consistency. Earlier, Bregler et al.~\cite{bregler2002turning} uses motion capture data to interpolate between user-defined keyframes, significantly reducing the time-intensive task of hand-drawing every frame. On similar lines, approaches like interactive contour tracking~\cite{agarwala2004keyframe} allow users to define object outlines that could be automatically interpolated across video frames. At the same time, bridging the 2D-3D gap,  Davis et al.~\cite{davis2006sketching} introduce a system that reconstructs 3D articulated figure poses from 2D sketches, allowing artists to predict keyframes. The system handles the ambiguity of 2D to 3D translation by providing the most likely pose and allowing for user refinement. It is further improved by Dalstein et al.~\cite{dalstein2015vector} that interpolates vector-based workflows by introducing topology-aware operations that could handle complex changes in shape and connectivity, supporting more robust and flexible animation of vector drawings. Further advancement comes with optimization and learning-based techniques to generate accurate, automated, and consistent sketch frames.

\begin{figure}[ht]
    \centering
    \includegraphics[width=\linewidth]{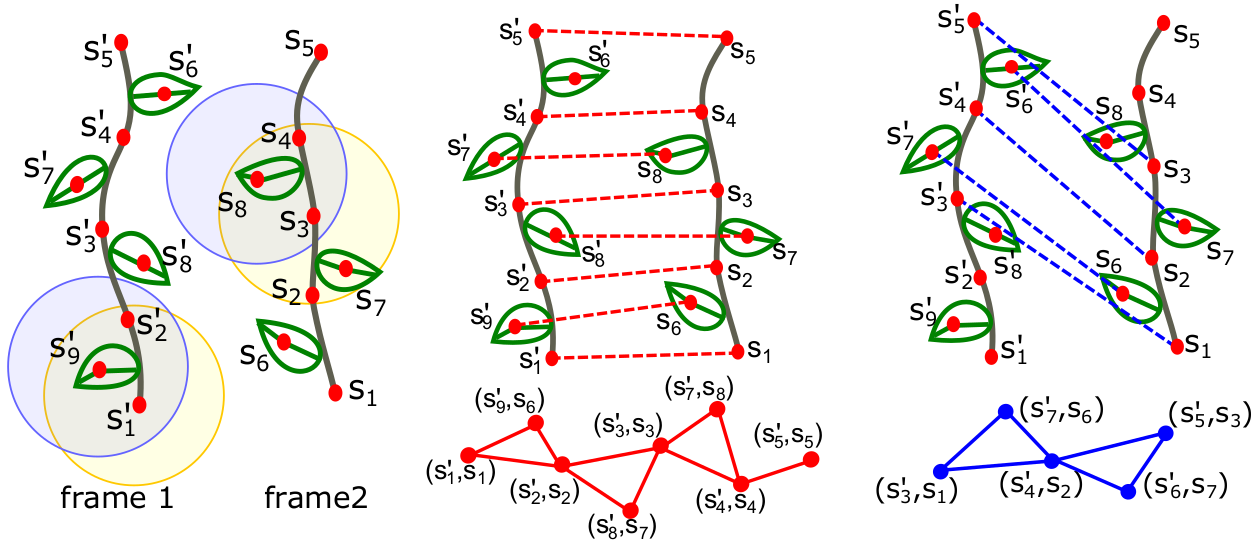} 
    \caption{Algorithm overview: as the user draws strokes in two frames, each new sample in Frame 2 is matched to Frame 1 by analyzing local and neighborhood similarity. Matching pairs are grouped into graphs representing consistent correspondences across frames. These graphs help refine matching over time, allowing the system to deform Frame 1 into Frame 2 as user guidance, shown as suggestions updated after each new stroke. Figure from~\cite{xing2015autocomplete}.}
    \label{fig:autocomplete}
\end{figure}

\subsection{Optimization-based methods}
To handle complex sketches, techniques emerge to tackle contour correspondence through patch alignment and clustering strategies to maintain consistent relationships between parts of complex objects~\cite{yu2011complex}. Other directions focused on animating static line drawings~\cite{fu2011animated, yang2012structure}, such as reconstructing plausible stroke orders for line illustrations, enabling temporally consistent animation, and applying structure-preserving deformation for articulated multi-part objects. BetweenIT~\cite{whited2010betweenit} advances through semi-automated stroke interpolation techniques that modeled trajectories with logarithmic spirals and curvature warps, focusing on generating tightly aligned inbetweens. Additionally, underlining classical animation principles, Guay et al.~\cite{guay2013line} incorporate the line of action as a structural and expressive guide for pose creation, aligning computational techniques with foundational animation principles. This trajectory highlights a progression from data-assisted interpolation to semantically aware, style-preserving stroke animation, reflecting an effort to blend automation with artistic intent in modern animation pipelines. Further, Quasi-3D rotation~\cite{furusawa2014quasi} enables interpolation of facial sketches across turning views by modeling approximate 3D motion from 2D inputs, allowing for convincing face rotations without requiring full 3D reconstruction. On similar lines, Autocomplete~\cite{xing2015autocomplete} presents a user-guided predictive system where the animator provides a keyframe stroke, and the system uses a graph-based pattern estimation module to predict the next stroke in the sequence, as shown in Fig.~\ref{fig:autocomplete}. This interactive approach allows users to accept, reject, or edit each predicted frame, balancing automation with creative oversight. These developments reflect a growing focus on structure-aware, predictive, and editable animation systems that accelerate the inbetweening process while respecting the user’s stylistic and compositional intent.  Recently, inbetweening techniques have progressed toward context-aware and structure-preserving approaches that enhance automation and artistic control. Initial improvements present contextual coherence via context-aware inbetweening~\cite{yang2017context}, which considers the spatial neighborhood of strokes to produce smoother and more visually consistent transitions. At the same time, DiLight~\cite{carvalho2017dilight} incorporates guideline-assisted curve correspondence, improving control over curved stroke behavior during interpolation.
At the structural level, adaptations of optical flow for line art introduce pixel-free motion estimation via distance transforms, making flow-based techniques applicable to sketch inputs~\cite{narita2019optical}. 

\begin{figure}[tbp]
    \centering
    \includegraphics[width=\linewidth]{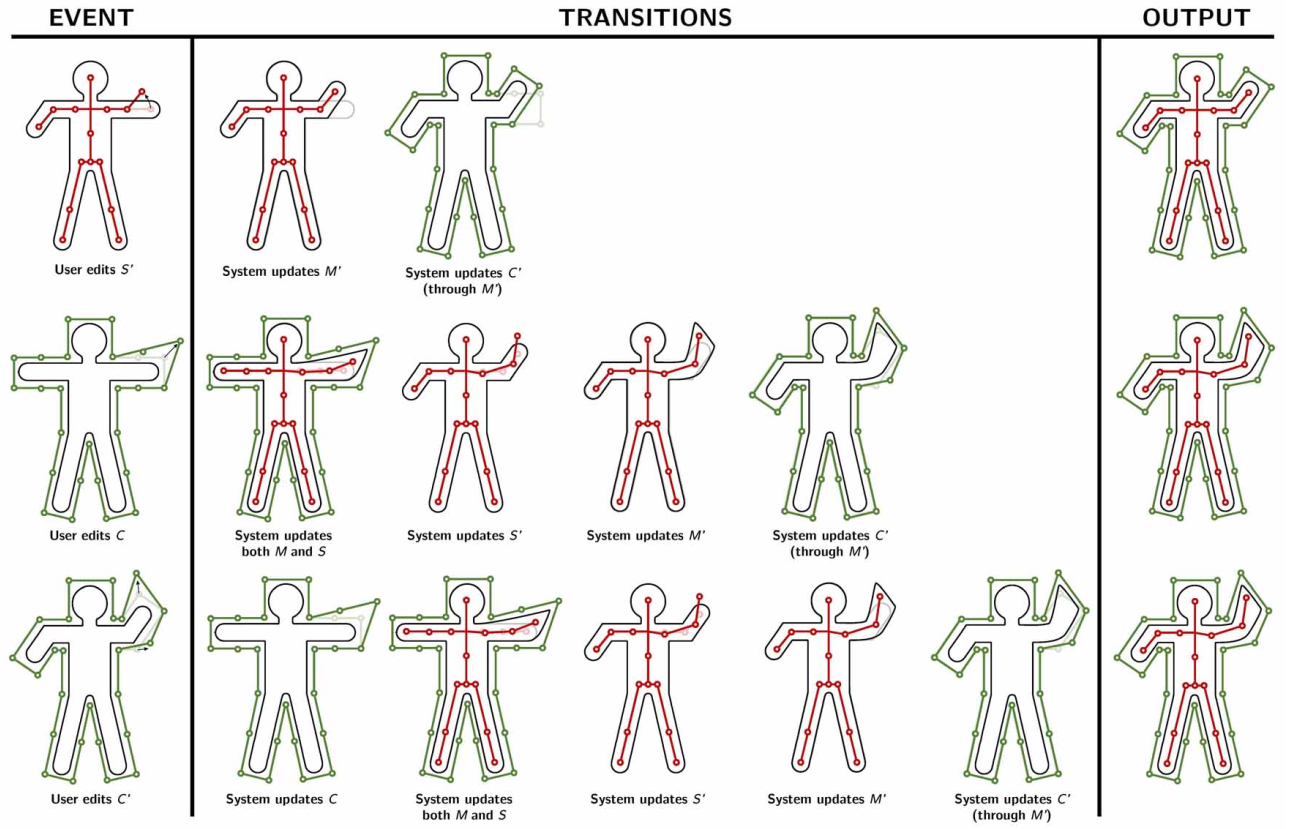} 
    \caption{The hybrid deformation framework enables real-time editing of either the skeleton in the deformed pose or the deformation cage in the rest or deformed configurations. All edits are automatically propagated to ensure consistent synchronization among the deformation components. Figure from~\cite{corda2020real}.}
    \label{fig:deformation}
\end{figure}

Several methods leverage artists' realistic drawing skills to create intuitive, artist-friendly controls that simplify complex tasks. The objective is to move from manual, low-level manipulation to high-level, artistic control that maintains visual fidelity and control. Hahn et al.~\cite{hahn2015sketch} introduce a sketch-based approach that enables intuitive static character posing. Extending from static to dynamic control, SketchiMo~\cite{choi2016sketchimo} introduces a sketch space that lets users edit motion trajectories and abstract animation constraints directly through drawing, with the system translating these sketches into optimization constraints for motion refinement. Simultaneously, hybrid deformation systems, Corda et al.~\cite{corda2020real}, enable unified manipulation of the posed skeleton, the rest cage, and the deformed cage. In this formulation, the articulated skeleton pose drives structural motion, while the rest cage defines the geometric embedding of the mesh. The resulting deformed cage $C'$, updated through coupling constraints, refines and stabilizes the surface deformation. By tightly synchronizing and with respect to, the framework combines the structural advantages of skeleton-driven articulation with the expressive flexibility of cage-based mesh deformation, as shown in Fig.~\ref{fig:deformation}.
These techniques established a robust theoretical and computational foundation for high-quality, real-time, and structure-preserving deformations, operating as core tools in modern sketch and character animation systems. 

\begin{figure}[ht]
    \centering
    \includegraphics[width=\linewidth]{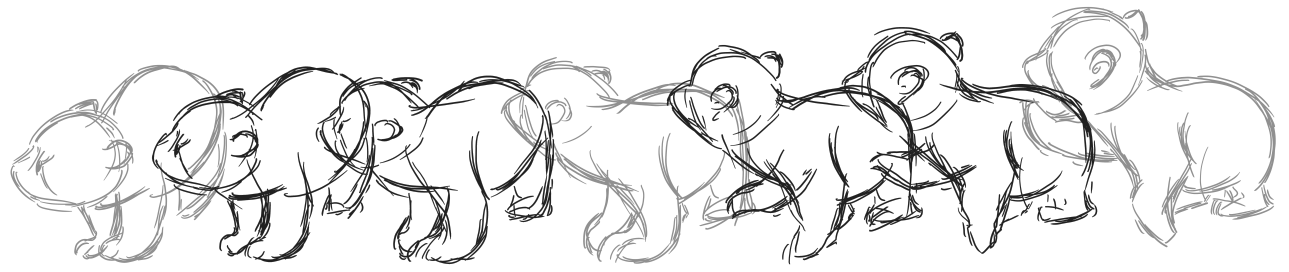} 
    \caption{Rough line-art interpolation. It takes rough vector key drawings (light gray), generates stylistically consistent intermediate strokes (black), and uses an interactive stroke distribution synthesis algorithm that minimizes temporal artifacts. Figure from~\cite{chen2023efficient}.}
    \label{fig:rough-art-interp}
\end{figure}

More advanced interpolation techniques have expanded the expressiveness and control in animation by incorporating structural awareness, partial automation, and user-in-the-loop prediction.  It is further extended by Shen et al.~\cite{shen2024bridging}, introducing multi-level sketch-aware interpolation with region, stroke, and pixel-level guidance, which preserves high-level structure and fine-grained detail. Later, to handle occlusions and depth in bitmap sketches, Brodt and Bessmeltsev~\cite{brodt2024skeleton} combine skeleton motion, bitmap deformation, and 2.5D lifting, enabling plausible interpolation even in complex scenes. Pushing toward rough-sketch workflows, the non-linear embedding-based method~\cite{even2023non} enables transient stroke representations that support expressive and non-deterministic interpolation for rough, exploratory sketches (see Fig.~\ref{fig:rough-art-interp}). Pushing further into rough sketch domains, Even et al.~\cite{even2025inbetweening} introduce occlusion-aware inbetweening for rough drawings with dynamic layouts and stroke visibility control to handle appearance in layered, evolving compositions. Recently, Cartoonimator~\cite{ranjan2025cartoonimator} introduce a paper-based tangible interface for keyframe animation, allowing users to create animations through physical manipulation of drawings rather than conventional digital timelines. The approach emphasizes embodied interaction to reduce the learning curve for novice animators and support creative exploration. The work highlights the potential of tangible and sketch-based systems to improve accessibility and engagement in animation authoring tools.

\begin{figure}[ht]
    \centering
    \includegraphics[width=\linewidth]{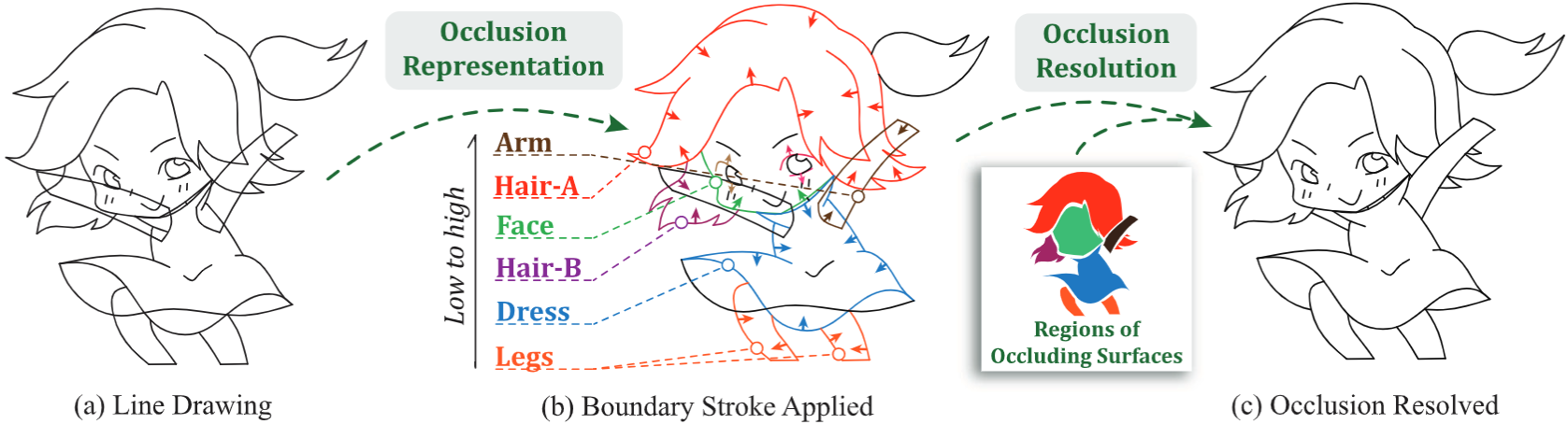} 
    \caption{Two-stage occlusion resolution in stroke-based drawings. (a) Original line drawing composed of multiple strokes with occlusions. (b) Occlusion representation using labeled boundary strokes, where half-arrows indicate relative depth ordering. (c) Occlusion resolution through identification of regions corresponding to occluding surfaces. Figure from~\cite{jiang2022stroke}.}
    \label{fig:inbetween}
\end{figure}

\subsection{Learning based interpolation techniques}
The advent of learning-based inbetweening brought structure awareness and generative capabilities to the field. Chen et al.~\cite{chen2023efficient} introduce a method for efficient stroke selection in vector-based interpolation to enable frame prediction, while self-occlusion is tackled through boundary-aware stroke propagation~\cite{jiang2022stroke} as given in Fig.~\ref{fig:inbetween}. These methods signal a shift toward flexible, semantically grounded, and structurally robust inbetweening frameworks capable of supporting a wide range of animation styles and input fidelity levels, from clean vector drawings to rough, gesture-based input.
Building on the evolution of inbetweening, recent methods introduce multi-level stroke correspondence, hybrid representations, and occlusion-aware deformation techniques to push the limits of automation and expressiveness in sketch-based animation. Siyao et al.~\cite{siyao2023deep} propose AnimeInbet, reframing line inbetweening as graph fusion of vertices. Mo et al.~\cite{mo2024joint} propose a method that enables accurate vector-stroke correspondence across frames, thereby improving temporal consistency in vector animations. Complementing this, Zhu et al.~\cite{zhu2025thin} introduce thin-plate spline-based interpolation with a keypoint-matching module, estimating per-frame keypoint flows and enabling smooth, large-motion interpolation between line drawings. These advancements reflect a trajectory toward semantically controlled, structure-aware, and occlusion-resilient inbetweening, making the animation of vector and exploratory rough drawings more accessible, flexible, and robust.

The trajectory reflects a progression: from geometry and contour-based methods to structure-preserving interpolation, then to learning-based, stroke- and skeleton-centric systems. Each stage addressed weaknesses of the previous methods, moving from simple contour tracking to handling complex multi-contour objects, then to occlusions, topology changes, and finally to deep models that capture stroke-level and semantic priors. Modern systems integrate multi-level correspondence (region, stroke, pixel), skeleton guidance, and deep generative priors to balance automation with artistic control. Combined, these approaches mark the transformation of inbetweening from a labor-intensive manual process into a semi-automated, structure- and style-preserving workflow, significantly reducing production costs while enlarging creative freedom.

\section{Physics-based techniques}
\label{sec:procedural}
\begin{figure}[ht]
    \centering
    \includegraphics[width=\linewidth]{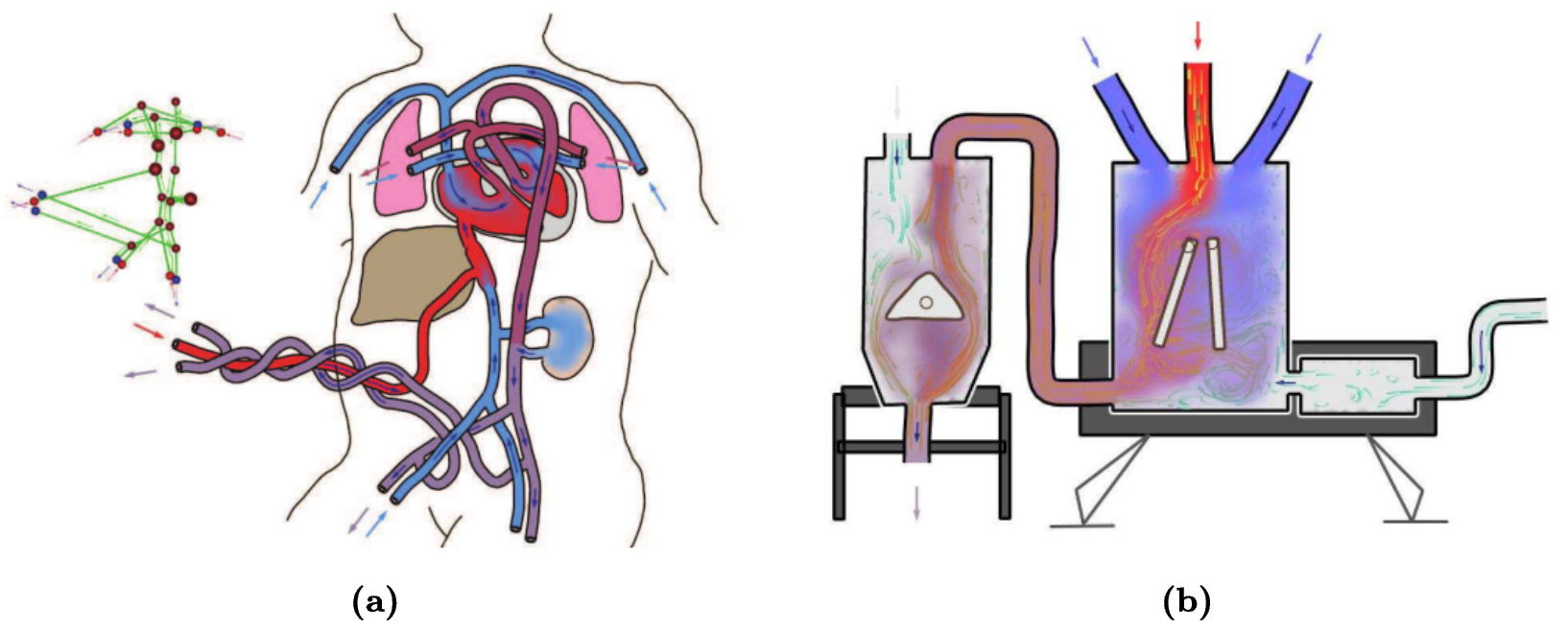} 
    \caption{Fluid systems in (a) physiology (fetal circulation) and (b) engineering (humidified agitator). Figure from~\cite{zhu2011sketch}.}
    \label{fig:fluid}
\end{figure}

In recent years, physics-based animation techniques have developed from early static images to immersive, real-world interactive storytelling, focusing on fluid simulation, physical constraints, and realistic motion. Traditional methods~\cite{chuang2005animating} use spectral noise synthesis to create stochastic motion textures for layered still images, producing looping video textures of passive elements. It is extended to realistic fluid motion by decomposing example videos into average images, flow fields, and residuals, allowing users to paint and refine custom flow patterns for target still images~\cite{okabe2009animating}. These approaches gradually evolved into more interactive systems, such as sketch-based hydraulic simulations and direct manipulation tools. These enabled interactive fluid illustration through 2.5D sketch-based hydraulic graphs~\cite{zhu2011sketch} with real-time simulation. This framework illustrates many fluid systems, such as physiological and engineering fluid systems, as shown in Fig.~\ref{fig:fluid}, while Physink~\cite{scott2013physink} brought direct manipulation of physics-enabled sketches for causal, editable animations. This evolution was advanced by tools that simplified novice workflows using cutout and layered interfaces~\cite{5601663}, and by system that introduces physically-based elastic motion into character rigs, enhancing expressive dynamics~\cite{guay2015adding}. Later, Lingen et al.~\cite{lingens2020towards} automate character animation through neural networks and evolutionary search in a physics-based environment. Integrating neural networks and generative models enables more refined control over fluid motion, including the generation of velocity fields from sketches~\cite{hu2019sketch2vf}. 
Immersive platforms expand this paradigm into AR, VR, and mixed reality, allowing animated sketches to respond to real-world objects for responsive motion, user gestures, and integrating motion and audio brushes for mixed-reality animated storyboards~\cite{eroglu2018fluid, suzuki2020realitysketch, li2023enchantedbrush}. 
Physics-based sketch animation methods lie in their progressive integration of procedural motion, user interactivity, and physical authenticity. Spectral and decomposition methods initially automated motion from static media, while later sketch-based and learning-based systems propose direct manipulation, physics constraints, and generative models for controllable, stylized motion. Immersive and AR/MR tools extended these concepts into real-world contexts, allowing dynamic sketches to respond to physical objects and environments. However, while these systems advanced artistic control and responsiveness, they still struggle to fully replicate realistic, physics-based fluid behavior over sketches, leaving a gap for future research in combining sketch-specific stylization with physically accurate fluid simulation.

\section{Data-driven techniques}
\label{sec:datadriven}
This section discusses data-driven techniques for sketch animation. In recent years, researchers have explored a variety of motion sources, including motion capture (MoCap) data, videos, and text, to extract motion information and transfer it into sketch-based representations.

\subsection{MoCap-based sketch animation methods}
MoCap offers a wide range of motion-driven, skeleton-based, and data-driven animation techniques to enhance motion consistency, stylization, and accessibility for experienced animators and beginners alike. Previously, efforts in this field focused on enhancing raw MoCap data with stylized 2D deformations, injecting artistic expressiveness into rigid 3D motion sequences~\cite{li2003stylizing}. Subsequent methods align MoCap trajectories with human body models~\cite{davis2006sketching}, producing accurate, realistic animations, although such approaches are limited in handling non-human or arbitrary forms. To introduce creative flexibility and secondary motion, systems emerge~\cite{li2006motionmaster, kim2008drivenshape} to retrieve and refine 3D motions from user-drawn pose sketches, enabling intuitive animation prototyping without complete MoCap setups and uses secondary motion that allows deformation of surrounding objects based on the dynamics of a driving shape. Other techniques support user-performed motion transfer in cutout-style animations and introduce skeleton-based deformation pipelines that simplify 2D animation production~\cite{barnes2008video, pan2011sketch}. Additionally, to bridge the physical correctness, refinements such as spacetime vertex constraints refined the MoCap sequences to meet physical plausibility and environmental constraints. It illustrates structured motion data, user input, and stylized control, paving the way for accessible yet physically grounded animation systems.

Recent methods have moved toward stroke-centric learning-based models. These methods are a progressive blending of motion capture, data-driven, sketch-based input, and learning-based representations to support both precision-driven animation (MoCap, skeleton embeddings, physical models) and creative, stylized, or accessible workflows (sketch-to-motion, style transfer, pose mapping). Further expanding the data-driven animation pipeline, techniques have prioritized accessibility, physical accuracy, and real-time interactivity, especially for users with minimal artistic or technical expertise. 3A2A~\cite{dadfar20213a2a} allows stick figure inputs to be transformed into expressive hand-drawn animations. Stroke-centric and sketch-guided techniques, such as those enabling motion transfer from video or example-based style propagation, now offer powerful tools for direct, intuitive control while embedding data-driven intelligence for automated motion inference. In sketch-based workflow, Patel et al.~\cite{patel2016tracemove} introduce Tracemove, a data-assisted system that offers frame-by-frame animation support by incorporating predefined motion patterns. While such tools offer greater control and stylistic guidance, they often rely on manual input across multiple stages, underlining the need for future systems to blend automation with intuitive authoring interfaces. Parallelly, data-driven systems focus on stylized rendering in line drawing animations from videos~\cite{ben2016line}.

Subsequent advancements introduce structure-aware deformation, skeleton-agnostic representations, and style-aware motion synthesis, expanding applicability to various motion domains and animation. In this direction, Pose2Pose~\cite{willett2020pose2pose} maps performer-specific gestures to 2D characters. Earlier methods used MoCap for motion retargeting but were limited to human-like motions; subsequent techniques added structure-aware deformation, skeleton-agnostic embeddings, and style-aware synthesis, enabling adaptation to diverse characters, styles, and motion domains. Further, ToonSynth~\cite{dvorovznak2018toonsynth} explores the technique of transferring style and motion by learning from hand-colored animations, enabling consistent stylistic and motion transfer to new skeletal sequences. Meanwhile, high-fidelity motion capture systems~\cite {kitagawa2020mocap} demonstrate their value in open-domain animation tasks, offering precise control and dense tracking, though with increased complexity and domain-specific constraints. These systems collectively reflect a broader push toward generalizable, stylized, and artifact-free animation pipelines that support creative flexibility and structural adaptability.
\begin{figure*}[ht]
    \centering
    \includegraphics[width=\linewidth]{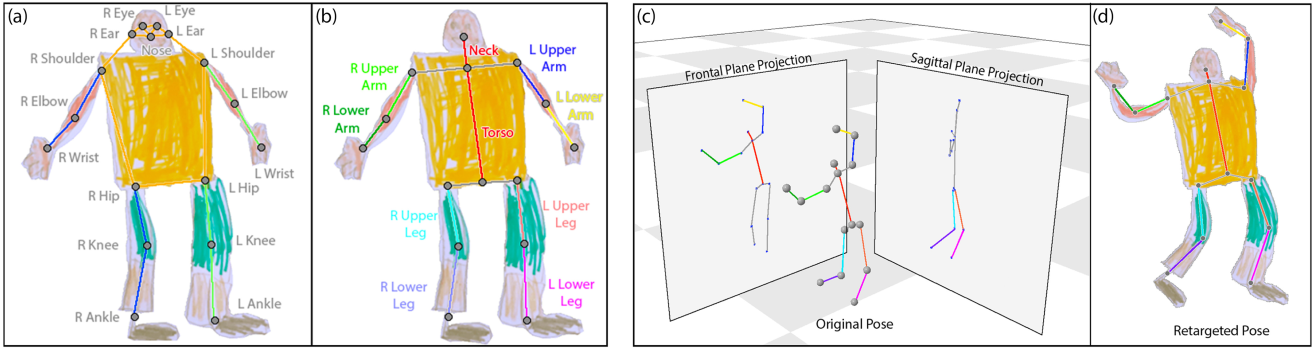} 
    \caption{This method constructs a skeletal rig from predicted joint keypoints (a) to animate the character (b). Motion is retargeted by projecting upper and lower body joints onto frontal and sagittal planes (c), aligning bone orientations to produce the final pose (d). Figure from~\cite{smith2023method}.}
    \label{fig:animation_drawing}
\end{figure*}

Meanwhile, Animated Drawings~\cite{smith2023method} integrates pose estimation from video, such as alphapose, with skeletal rigging to animate hand-drawn characters. An ARAP deformation loss preserves structural consistency during joint rotations, maintaining the visual fidelity of the original sketch given in Fig.~\ref{fig:animation_drawing}. This system constructs a character rig by projecting motion-capture anchors to skeletal joints and computing bone orientations within their respective planes. However, this approach is primarily limited to bipedal motion, reducing its generalizability to non-human objects or characters. These stroke-level and pose-to-sketch systems represent a promising step toward precise, temporally-aware, and artist-preserving animation frameworks, balancing data-driven automation and user-controllable semantics in sketch-based animation. 

\subsection{Video-driven sketch animation}
Traditional sketch animation tools primarily rely on keyframe-based animation or motion capture (MoCap) data. Video motion retargeting aims to map the video motion from source to target, which enables identity-preserving motion and controllable motion transfer across different identities. Cross-domain video motion retargeting addresses the challenge of motion transfer between different domains, such as real video to sketch, avatars, and cartoon characters. Traditional motion retargeting tools operate within a single domain and do not tackle the domain gap in appearance, articulation, and temporal coherence. Single-domain motion models (SDMM)~\cite{siarohin2019first, siarohin2021motion, tao2022motion} are not suitable for sketches due to the distinctive nature of sketches. Cross-domain motion models~\cite{wang2020self, xu2022motion, zhao2024pose, hu2024animate} overcome the challenges of SDMM by establishing the pose correspondence and alignment. However, sketch animation remains a challenge, and these methods cannot preserve the appearance and motion information. Further, a cross-domain motion model JOKR~\cite{mokady2022jokr} proposes a framework that learns joint keypoints of input videos and domain confusion loss for shared representation. Further, the shared feature information is passed to the generator module to generate animated sequences. The advantage of this method is that it does not require a large dataset for training. This method represents significant strides in bridging the gap of cross-the-motion retargeting and tackling identity preservation, stylization, and temporal consistency challenges. Further, cross-domain video motion retargeting for sketch animation, moving from early pose-alignment and keypoint-based mapping to modern diffusion- and stroke-level approaches that enable automatic, flexible, and controllable motion transfer while preserving sketch appearance and structure. Diffusion-based methods~\cite{xie2024video, xu2024magicanimate, hu2025animate} require extensive training data and provide slow inference. Also, it does not give attention to motion consistency in the animated sketches. 
\begin{figure}[ht]
    \centering
    \includegraphics[width=\linewidth]{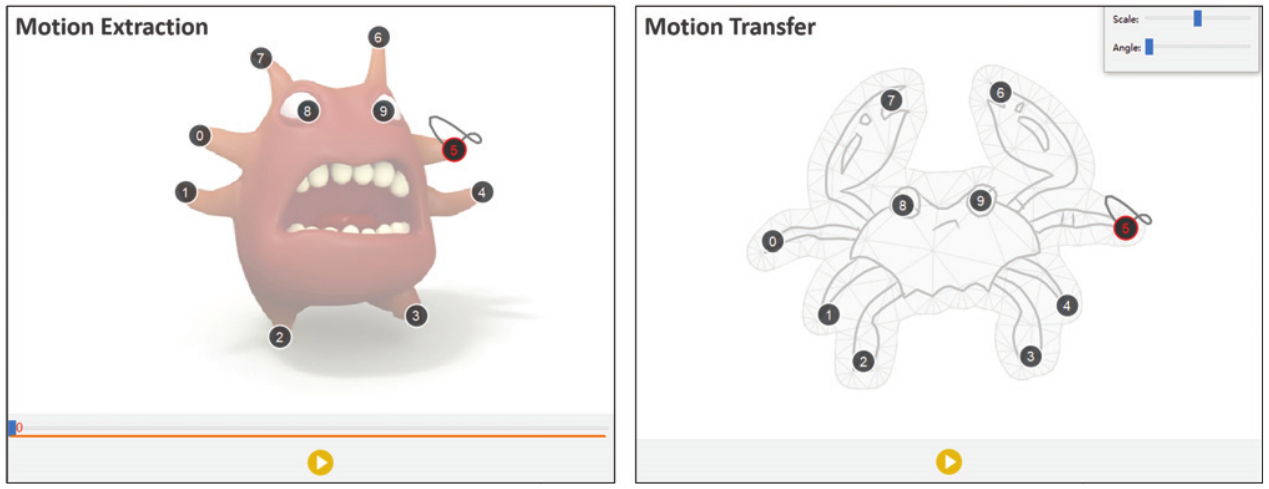} 
    \caption{Motion extraction and transfer workflow. Figure from~\cite{su2018live}}
    \label{fig:live_model}
\end{figure}

\begin{figure}[ht]
    \centering
    \includegraphics[width=\linewidth]{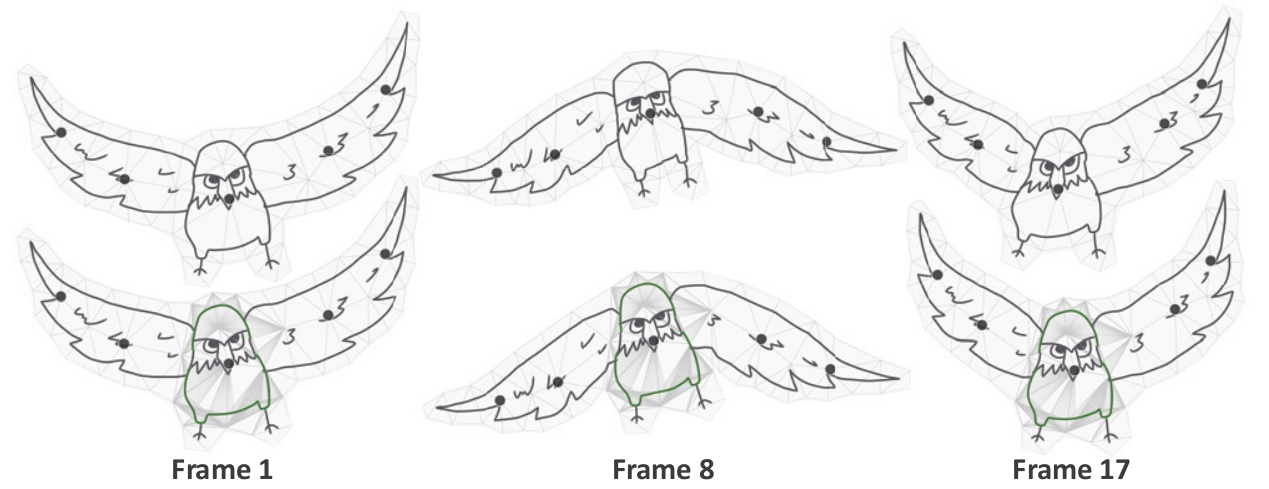} 
    \caption{Stroke preserving as-rigid-as-possible. Figure from~\cite{su2018live}}
    \label{fig:arap_loss}
\end{figure}

With the rise of video-assisted sketch animation, researchers have focused on extracting motion cues from real-world videos and transferring them to freehand sketches, enabling dynamic animation from static drawings. Live Sketch~\cite{su2018live} introduces a video-driven sketch animation framework by presenting a control point-based tracking technique that estimates motion trajectories from video and maps them onto corresponding sketch control points (see Fig.~\ref{fig:live_model}). This two-stage pipeline motion extraction and motion transfer allows video-based motion to drive the deformation and animation of sketches. To preserve the visual quality of drawn elements, the system uses a multi-layered approach that handles self-occlusion and a stroke-preserving ARAP function for rigidity constraints, ensuring the strokes maintain their original shape during deformation, as shown in Fig.~\ref{fig:arap_loss}. 

\begin{figure}[ht]
    \centering
    \includegraphics[width=\linewidth]{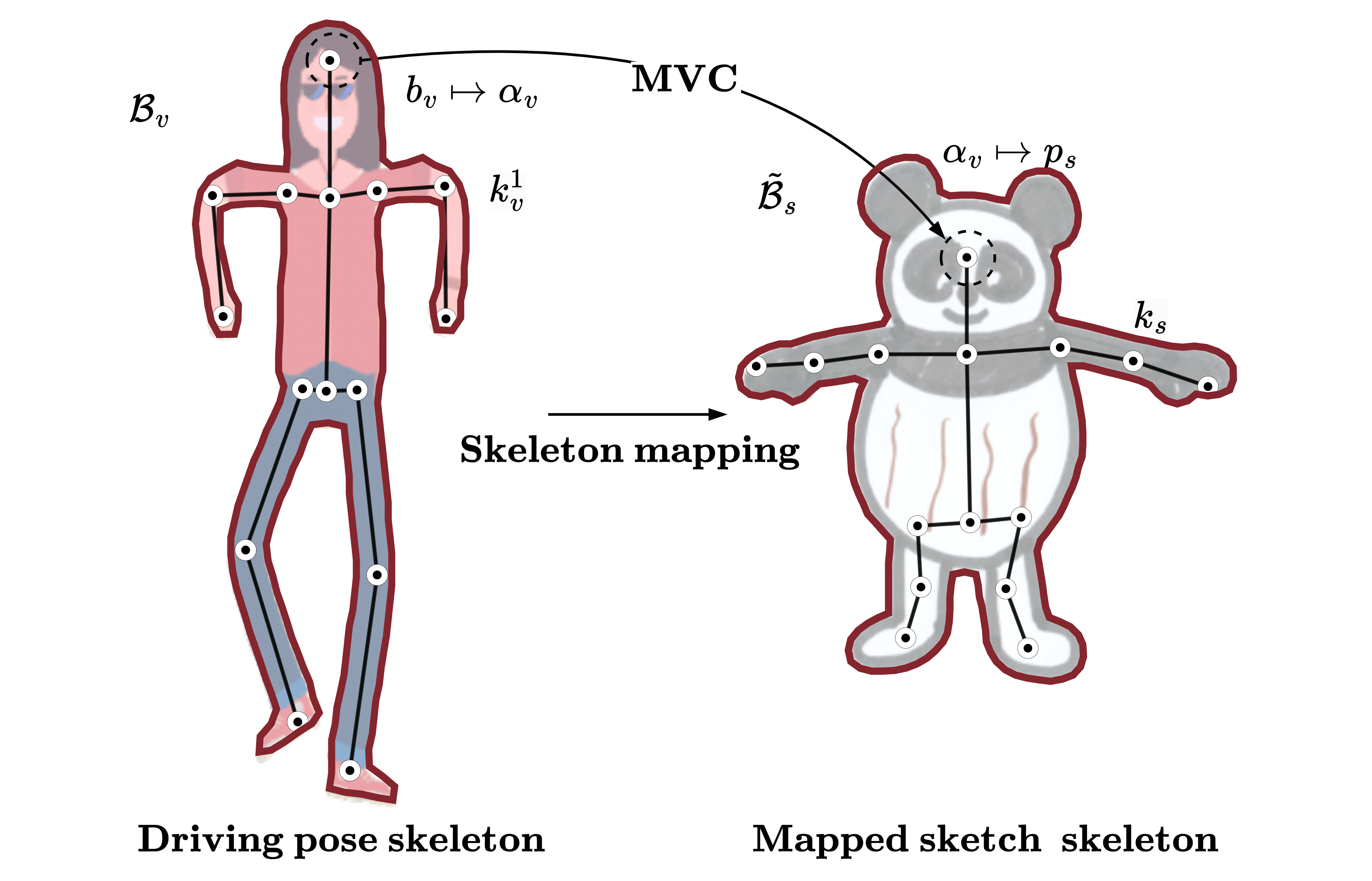} 
    \caption{Video to sketch skeleton mapping using mean value coordinates. Figure from~\cite{rai2024sketchanim}.}
    \label{fig:skeleton_mapping}
\end{figure}

Recent advances in sketch animation have increased focus on stroke-level representations, enabling better control over shape and motion while improving temporal consistency and interpretability. In this direction, Wakey-Wakey~\cite{xie2023wakey} proposes a stroke-level representation capable of generating sketch animation by disentangling shape and motion, allowing independent control of object geometry and motion, and producing temporally consistent animations from static input. Further, sketch animation methods have increasingly focused on video-driven motion customization and cross-domain compatibility, enabling more expressive and adaptable workflows. In this domain, SketchAnimator~\cite{yang2024sketchanimator} introduces a one-shot motion customization pipeline through a structured three-stage process: appearance learning, motion learning, and video prior distillation using SDS loss. This framework allows users to transfer motion from a reference video to a sketch while preserving visual style and temporal consistency. Expanding on this, Xie et al.~\cite{xie2024video} propose a cyclic reconstruction mechanism to improve motion consistency and domain adaptability. However, its reliance on domain-specific training data and limitation to specific motion types restricts generalization. 
Addressing these drawbacks, SketchAnim~\cite{rai2024sketchanim} introduces a video skeleton driving motion mapping module to animate the hand-drawn input sketch. It performs point tracking to extract skeletal motion, then uses shape-matching-based skeleton mapping between the video skeleton and the sketch using mean value coordinates~\cite{floater2003mean} for smooth deformation, as shown in Fig.~\ref{fig:skeleton_mapping}. Additionally, it incorporates a discrete depth value to handle the self-occlusion during sketch deformation. This method provides the flexibility of generating biped, quadruped, and inanimate sketch animation. Despite its effectiveness, SketchAnim remains constrained by its inability to handle 3D motions such as frontal views, and is sensitive to shape correspondence between the video skeleton and sketch layout. 

Recently, Zhu et al.~\cite{zhu2025vector} introduce a differentiable motion trajectory–based framework that models vector sketch stroke motion as continuous polynomial functions, enabling globally optimized, temporally coherent sketch animations with reduced flickering and improved consistency.

These works advanced integration of motion extraction, cross-domain adaptation, and appearance preservation techniques. Previous methods provided coarse motion transfer but lacked domain-specific fidelity; mid-stage approaches introduce stroke-level and region-guided constraints to preserve sketch appearance during deformation; recent methods integrate diffusion priors, motion decomposition, and skeleton-based mapping to achieve higher fidelity, temporal coherence, and flexibility across motion types. It marks a shift from domain-agnostic motion mapping toward sketch-aware, structure-preserving, and semantically controllable animation pipelines, significantly lowering the manual workload in 2D sketch animation production while expanding creative control. It highlights the growing focus on motion fidelity, adaptability, and semantic mapping in sketch animation pipelines, while highlighting open challenges such as 3D generalization, data efficiency, and occlusion handling in freeform animation systems.

Further, combining large language models and natural language understanding with sketch-based animation has evolved a new generation of systems where users can drive motion, expression, and interaction in sketches using simple text prompts. 

\subsection{Text-driven sketch animation}
Recently, a wave of research marks a decisive shift in sketch animation toward semantic, language-driven, and diffusion-powered workflows, enabling users to animate complex sketches, clipart, and story-driven scenes with minimal manual keyframing. It moves beyond traditional keyframing by leveraging text-to-video diffusion models and geometry-aware losses to animate complex scenes in a controllable yet automated way. Systems such as Gal et al.~\cite{gal2024breathing} pioneered this direction by introducing a text-to-sketch animation by using pretrained text-to-video diffusion models with score distillation sampling (SDS)~\cite{poole2022dreamfusion} loss to animate Bézier-curve sketches using local deformation and global transformation modules. Building on this, the follow-up method~\cite{rai2024enhancing} introduces length–area regularization and ARAP-based rigidity constraints to improve temporal smoothness and preserve stroke structure during animation. Further advancement adapts diffusion priors for flipbook-style raster animations~\cite{bandyopadhyay2025flipsketch}, introducing fine-tuning for sketch-style frames, reference frame noise refinement, and dual-attention composition for visual consistency. These approaches combine static art and temporal storytelling with minimal artist effort. Extending beyond sketches, Dynamic Typography~\cite{liu2025dynamic} advances text-prompt animation to deform and animate letters while preserving legibility, and AniClipart~\cite{wu2025aniclipart} applies Bézier-curve motion regularization and ARAP deformation to clipart, followed by FlexiClip~\cite{khandelwal2025flexiclip}, which adds temporal Jacobians, continuous-time modeling, and flow matching loss for smooth, coherent clipart animations. These methods redefine the sketch animation pipeline by tightly integrating semantic control, diffusion generation, and geometric regularization, setting a foundation for accessible yet expressive animation.

\begin{figure}[ht]
    \centering
    \includegraphics[width=0.6\linewidth]{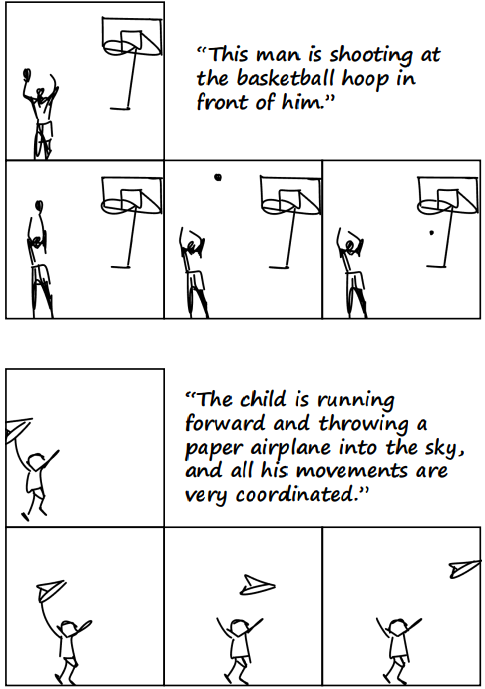} 
    \caption{Overview of a text-driven sketch animation framework. The model takes a vector sketch and a motion prompt as input and generates a short video in which the sketch is animated according to the specified motion description. Figure from~\cite{liang2025multi}.}
    \label{fig:multi-sketch}
\end{figure}

Recognizing the limitations of earlier sketch animation methods, recent approaches have significantly advanced toward multi-object, instance-aware, and story-driven animation. These techniques enable the animation of complex vector scenes while maintaining semantic coherence, motion synchronization, and storytelling. For instance, AnimateSketches~\cite{deng2025animatesketches} introduces prompt-guided instance-aware masks (PGIM) and mask-based SDS to animate multi-object vector sketches without affecting the rest of the scene elements, while MoSketch~\cite{liu2025multi} combined LLM-based scene parsing and motion decomposition with compositional SDS loss to achieve synchronized multi-object animation. Further, Liang et al.~\cite{liang2025multi} propose a structured two-stage pipeline for animating complex vector sketches with multiple objects, consisting of motion initialization and motion refinement, where users can group sketch parts and assign keyframes. A group-based displacement network (GDN), enhanced with context-conditioned features and a text-to-video diffusion backbone, refined these motions to produce temporally consistent, smooth animations, making the workflow intuitive, as shown in Fig.~\ref{fig:multi-sketch}. 
Stepping into narrative generation, FairyGen~\cite{zheng2025fairygen} automates story-driven cartoon creation from a single child’s drawing by using MLLM-based storyboarding, style-propagation adapters, cinematic shot design, 3D proxy motion reconstruction, and an image-to-video diffusion model with a two-stage motion customization adapter for identity preservation and temporal modeling. These methods mark a substantial hop toward semantic, multi-object, and story-centric animation systems, blending LLM reasoning, visual grounding, and diffusion-powered generation to achieve expressive, scalable, and accessible animated content creation.

These works have a clear trajectory from stroke-level motion control to text-driven scene animation using large pretrained diffusion and language models. Previous methods and their enhancements tackled single-object, motion-preserving animation, which evolved into style-preserving and temporally consistent pipelines. Recent advances address multi-object awareness and complex motion representation, culminating in full-scene, story-driven systems that integrate LLM-based planning, style transfer, and physically plausible motion. It marks a paradigm shift toward intuitive, text-and-sketch-driven authoring pipelines in which artistic style, temporal coherence, and narrative structure are jointly optimized, drastically lowering the barrier to creating high-quality, expressive animations.

\section{2D sketch to 3D animation}
\label{sec:2d3d}
2D-to-3D character animation has emerged as an effective research area that aims to transform static 2D sketches into animated 3D models. Initial method~\cite{reinert2016animated} focuses on generating 3D textured animation sequences from the 2D video using skeletal sketching, followed by segmentation, tracking, and texturing. Building on this foundation, Bessmeltsev et al.~\cite{bessmeltsev2016gesture3d} introduce gesture-based sketching on input character models, projecting 2D skeletons to estimate new poses. It allows for more expressive user input but still relies heavily on existing 3D templates, limiting the generation of entirely novel characters. Later systems, such as MagicToon~\cite{feng2017magictoon}, integrate an augmented reality interface in which the user draws a 2D sketch, creates a 3D model, and deforms it using skeletal deformation. It provides the accessibility of a novice user to generate a 2D sketch to a 3D model for manipulation and rigging. 

Building upon previous sketch-to-3D animation frameworks, Weng et al.~\cite{weng2019photo} propose Photo Wake-up, which takes an image input, performs segmentation and 2D pose estimation, and uses an SMPL~\cite{loper2023smpl} template model to project the 2D pose on the image with a skinning map. It uses a rigged 3D mesh with a combined depth and skinning map to animate it with Mocap, and textures it with an inpainting background. Progressing further, sketch-based approach~\cite{brodt2022sketch2pose} enables direct 3D pose retrieval from bitmap inputs without requiring detailed structural information. The pose uses custom rigging and skinning using standard retargeting tools. It enabled freehand creative input but focused on pose reconstruction rather than full 3D textured animation. Although this improved motion consistency and flexibility, it requires a detailed sketch structure, and authenticity was not completely ensured. Simultaneously, a more playful and intuitive system called MonsterMash~\cite{dvorovzvnak2020monster} introduces an interactive solution where users draw the strokes, and the method inflates the sketch drawing and animation in real-time into a 3D deformable model. It proposes a user-friendly interface for 3D character creation from 2D drawing and manipulation by integrating sketch-based modeling with a physics-inspired animation system. This trajectory reflects a shift toward systems highlighting real-time interactivity, minimal user effort, and more prominent creative tools, bridging the gap between simple sketching and expressive 3D animation.

\begin{figure}[ht]
    \centering
    \includegraphics[width=\linewidth]{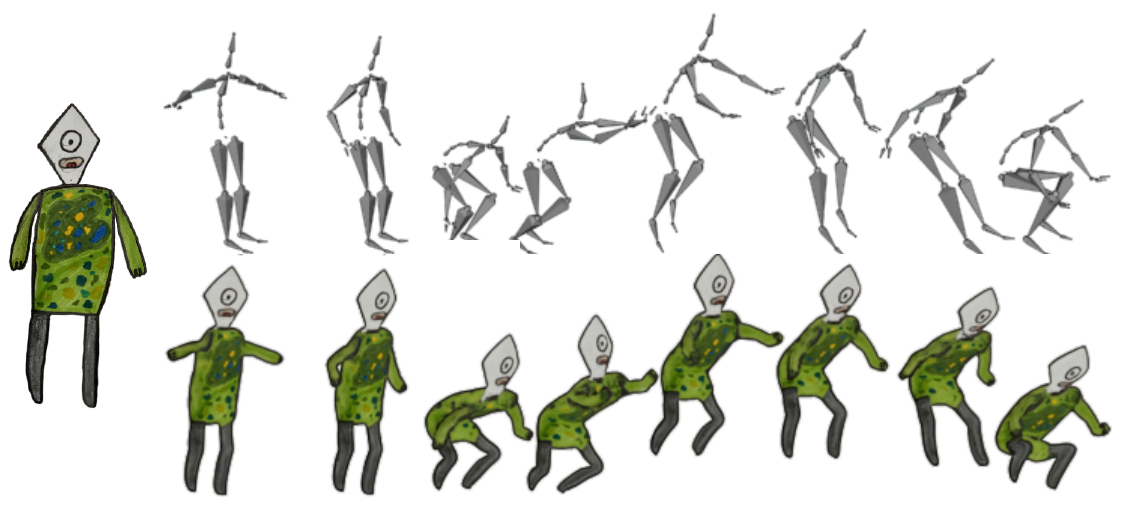} 
    \caption{DrawingSpinUp method generates a 2D sketch of a 3D model and animates it using the target motion. Figure from~\cite{zhou2024drawingspinup}.}
    \label{fig:spinup}
\end{figure}

In recent years, hand-drawn 2D sketch to 3D model generation techniques~\cite{zhang2022creatureshop, thiault2025spineloft, song2024magiccartoon, zhao2025hunyuan3d, li2025step1x} provide an edge to generate the textured animation-ready 3D meshes directly from 2D sketches. These methods provide high-fidelity reconstructions with rich detail and enable animation-ready meshes. However, they often depend on user-provided constraints or structural hints, limiting their utility for unstructured freehand art. Removing this dependency for the complex structure of 2D textured sketches, DrawingSpinUp~\cite{zhou2024drawingspinup} presents a framework to animate characters from single drawings by inferring 3D geometry and motion cues using Mixamo deformation priors and shape-consistency networks (see Fig.~\ref{fig:spinup}). It synthesizes plausible 3D motion without requiring complicated user input, thus bridging the gap between casual 2D sketching and dynamic 3D character animation, and opening up new possibilities for accessible, sketch-driven animation pipelines.

\begin{figure}[tbp]
    \centering
    \includegraphics[width=\linewidth]{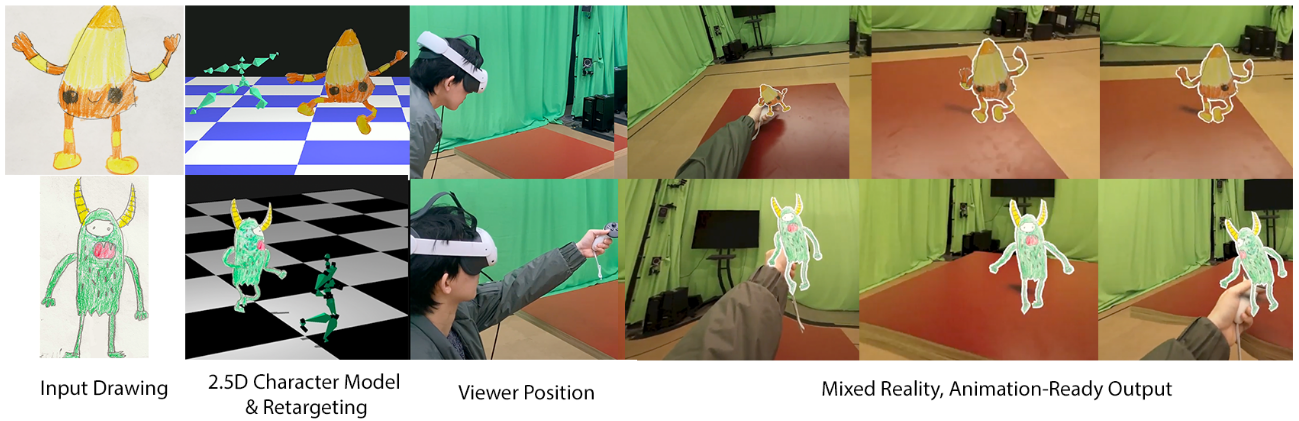} 
    \caption{A real-time animation system that transforms a single childlike drawing into a 2.5D character viewable from multiple angles. It supports motion retargeting using any 3D skeletal motion, making it essential for mixed-reality applications. Figure from~\cite{smith2025animating}.}
    \label{fig:25D_AnimDraw}
\end{figure}

Recent advancements in 2D-to-3D character animation have increasingly prioritized freeform sketch input, shape consistency, and robustness to occlusion, pushing the boundaries of accessibility and visual fidelity. It allows freeform sketch drawing to 3D deformation with shape consistency using stylization, without providing additional structural information. These models achieve consistent shape-preserving stylization across motion sequences, but face challenges such as self-occlusion and visual artifacts. Addressing this, Yoon et al.~\cite{yoon2025occlusion} introduce an occlusion-robust stylization technique that integrates flow-depth edge detection (FDED) to infer reliable edge maps even under unseen or occluded poses. This framework ensures smooth, stable, and non-jittery animated sketches even in the presence of self-occlusion. On a similar note, Zhou et al.~\cite{zhou2025rigging} introduce a hybrid animation approach that combines 3D skeletal guidance with diffusion-based video generation to produce natural, temporally coherent animations of hand-drawn 2D characters. By refining rigged motion with a domain-adapted diffusion model that adds realistic secondary dynamics, it produces expressive, high-quality animations while preserving the original artistic style.
Complementing this, Smith et al.,~\cite{smith2025animating} introduce an animation systems that takes a single childlike figure drawing and convert it into a 2.5D character model, which is animated using  3D skeletal motion and viewed from different views. It offers a Mixed-reality application suitable for modeling and animation in real-time and generates view-dependent motion retargeting as shown in Fig.~\ref{fig:25D_AnimDraw}.
Further, sketch-based character animation has reached a new milestone with the introduction of end-to-end systems capable of generating fully textured, rigged, and animated 3D characters directly from freehand sketches and narrative instructions~\cite{Zhong:2025:Sketch2Anim}. This approach represents a significant leap in expressiveness and automation, unifying storyboarding, rigging, and motion synthesis into a single pipeline. Integrating sketch-based priors with semantic action cues represents the most advanced step yet, enabling textured, rigged, and animated 3D characters directly from freehand sketches and narrative instructions. However, the computational cost of large-scale learning models and the challenge of balancing automation with fine-grained user control remain open research problems.

\section{Sketch animation interfaces}
\label{sec:interfaces}
Sketch-based animation has made a remarkable evolution in the past few years. Starting with the keyframe and data-driven methods to advance automatic and stylized techniques using generative models, researchers have constantly pushed the boundaries of how novice animators can create animation smoothly with less time and effort, without compromising the animation quality. Early systems in sketch-based animation focused on enhancing traditional workflows by integrating computational capabilities directly into the artistic process. One major direction involves augmenting hand-drawn keyframe animation~\cite{jain2010augmenting} with 3D physical effects such as cloth, fluids, and particles into 2D drawings, enabling secondary motion while preserving the animator’s control. Efforts have emerged to digitize tactile art forms through multitouch platforms that recreate sand animation, simplify creation, mix with video, and enable gesture-based replay, bridging the gap between tactile artistry and digital flexibility~\cite{kazi2011sandcanvas}. These developments emphasized preserving the expressiveness of handcrafted input while offering digital flexibility. 
Stylization tools~\cite{kazi2012vignette} extend this principle to pen-and-ink illustrations, enabling partial texture drawing with automatic synthesis and editable filling, preserving the artist’s style while reducing repetitive work. Sketch-based interfaces soon expanded into educational contexts~\cite{cheema2012physicsbook}, automatically animating sketched physics diagrams by recognizing shapes, annotations, and equations, and using a domain-specific physics engine for validation, illustrating how sketch-based interfaces can integrate reasoning and visualization. Later, systems tackled the challenge of bringing static drawings to life by allowing artists to define example poses, interpolation rules, and procedural physics-based transitions, thus merging handcrafted pose design with simulation-driven motion for broader reusability~\cite{jones2013dynamic}. 
\begin{figure}[ht]
    \centering
    \includegraphics[width=\linewidth]{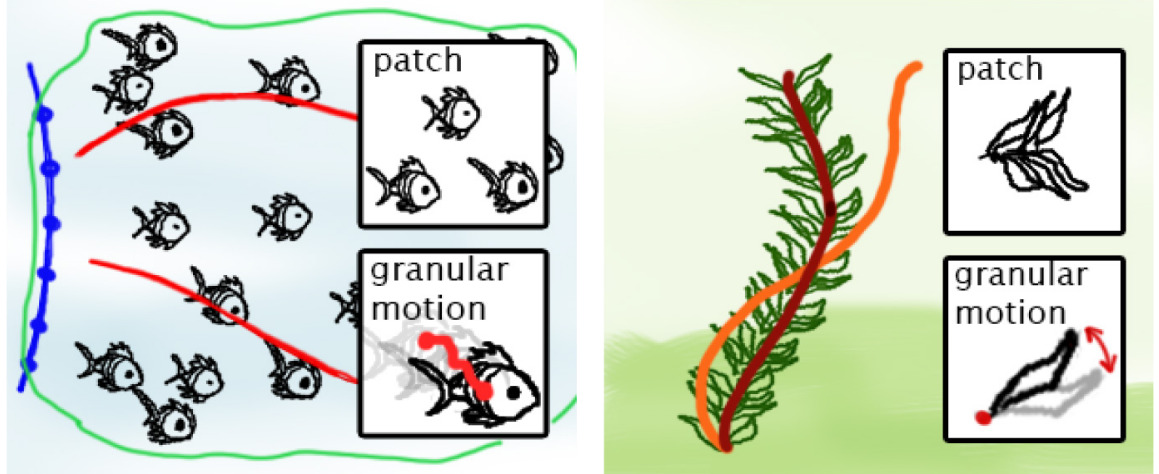} 
    \caption{Given two types of kinetic textures: (a) Emitting texture uses a source patch, emitter (blue), global paths (red), and granular motion. (b) Oscillating texture uses a brush skeleton (brown), an oscillating skeleton (orange), and granular motion. Figure from~\cite{kazi2014draco}.}
    \label{fig:draco}
\end{figure}
Further, researchers made a significant effort to introduce dynamic and expressive possibilities for animated illustrations. A wave of sketch-based animation tools~\cite{kazi2014draco, kazi2014kitty} that move beyond aiding static-to-dynamic transformation toward rich behavioral control, procedural effects, and real-time interaction, while maintaining an accessible interface for both novices and experts. In the continued effort to facilitate and enrich sketch-based animation, subsequent systems introduce more dynamic and interactive capabilities while preserving the expressive integrity of hand-drawn illustrations. One key advancement enables artists to embed continuous, coordinated motions into static sketches using motion handles and kinetic textures (see Fig.~\ref{fig:draco}), enabling dynamic environmental effects like swaying leaves or rippling water without disrupting the timeless quality of still art~\cite{kazi2014draco}. This approach maintained the timeless visual style while layering in dynamic elements.

\begin{figure}[ht]
    \centering
    \includegraphics[width=\linewidth]{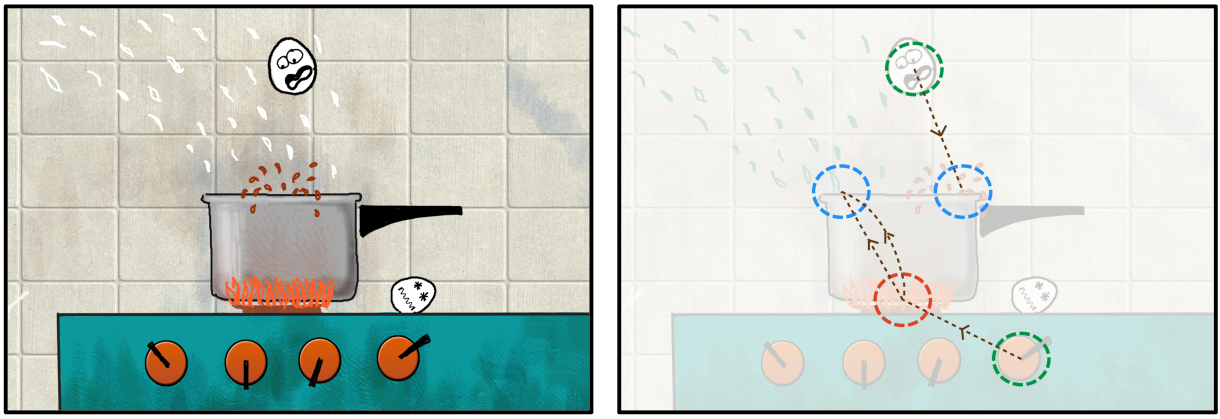} 
    \caption{Functional relationship graph of the given corresponding figure. Figure from~\cite{kazi2014kitty}.}
    \label{fig:kitty}
\end{figure}

\begin{figure}[tbp]
    \centering
    \includegraphics[width=\linewidth]{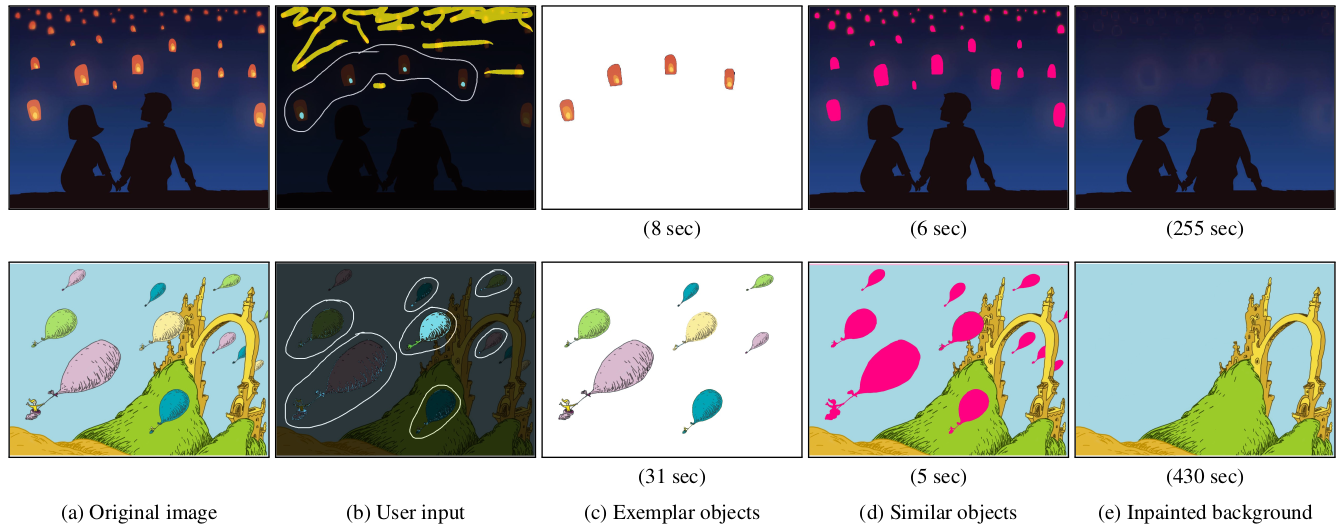} 
    \caption{Summary of segmentation pipeline: (a) original image, (b) user input (white: exemplars, green: foreground, yellow: selection hints), (c) system-detected exemplars, (d) dark pink: matched objects for inpainting, (e) final background layer. Figure from~\cite{willett2018mixed}.}
    \label{fig:mixed}
\end{figure}

Building on interactive storytelling, Kitty~\cite{kazi2014kitty} extended sketch-based animation to include functional relationships between entities via a graph-based model, as shown in Fig.~\ref{fig:kitty}, narrative interactivity, making them suitable for applications in infographics, educational content, and interactive storytelling such as children’s books, where visual elements could behave in contextually appropriate and responsive ways.
Sketch-based animation continued with systems that expanded the scope of motion control, expressiveness, and interactivity while minimizing dependency on complex inputs like motion capture or dense datasets. More advancements enable artists to define complex 3D motions from single strokes through space-time curves and dynamic lines of action~\cite{guay2015space}, offering complete geometric control without requiring motion capture or large datasets. Building on direct manipulation, adding user-controllable motion effects using brushes introduces user-controlled stylized effects called EnergyBrushes~\cite{xing2016energy}, which shifted toward stylized natural phenomena, allowing coarse energy strokes to drive particle-based simulations of water, fire, or smoke, combining physical plausibility with artistic expressiveness. Further simplification came through a modular exaggeration tool~\cite{kazi2016motion} that allows users to deform and enhance motion effects using plug-and-play amplifiers, enabling rapid iterations even by novices. Real-time performance contexts were also addressed, with interfaces that triggered animations via multitouch input and predictive suggestions~\cite{willett2017triggering}, and systems that automate the addition of physically plausible secondary motion~\cite{willett2017secondary} to live-performed animations using parameterized rigs. On similar lines, Ciccone et al.~\cite{ciccone2017authoring} targets cyclic animation creation, offering a capture-based interface to extract and edit loops easily, while a mixed-initiative system~\cite{willett_mixed-initiative_2018} emerges that progressively layered control and automation, allowing artists to animate still images using scribbles for segmentation, texture extraction, and kinetic texture synthesis, as shown in Fig.~\ref{fig:mixed} by bridging static composition with dynamic expressiveness while maintaining a high degree of user agency and creative flexibility.

\begin{figure}[tbp]
    \centering
    \includegraphics[width=\linewidth]{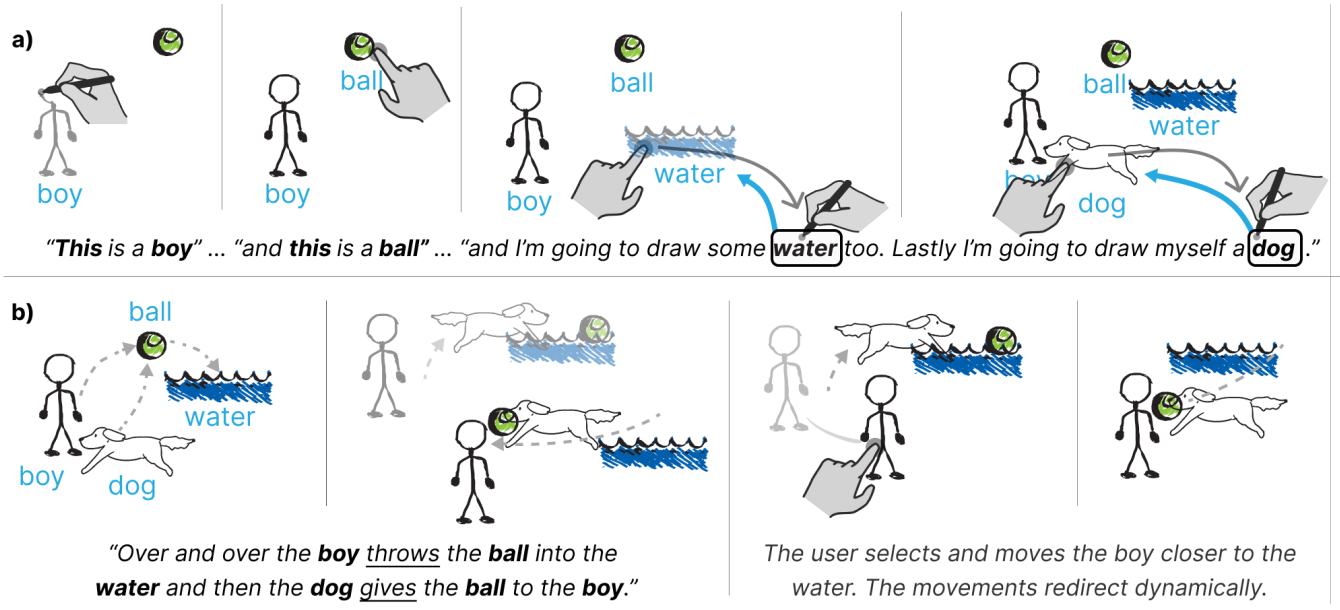} 
    \caption{Overall workflow: (a) The user draws and labels objects using multimodal input pen, touch, and speech at various stages: during drawing, after drawing, via transcript selection, or by removing labels. (b) Labeled sketches respond to commands; Right: the user interactively moves objects while the system dynamically adjusts simulated motion. Figure from~\cite{rosenberg2024drawtalking}.}
    \label{fig:anim_tool}
\end{figure}

In recent years, sketch-based and immersive animation tools have highlighted cross-modal interaction, immersive environments, and generative AI integration to expand creative possibilities beyond traditional 2D interfaces. For instance, augmented reality platform~\cite{xia2023realitycanvas} advances AR sketching by enabling responsive, improvisational animation in real-world contexts, offering six animation primitives such as object binding, particle effects, and flip-book animation derived from real-world scribble animation practices, making it suitable for storytelling, education, and prototyping. Parallelly, cross-modal systems push natural interaction by combining sketching and speech~\cite{rosenberg2024drawtalking} to create and control interactive worlds, giving users programming-like control without coding and enabling fluid, improvisational world-building for narrative contexts given in Fig.~\ref{fig:anim_tool}. In a production setting, Guajardo et al.~\cite{guajardo2024generative} demonstrate how generative models can integrate into professional workflows to facilitate the creative pipeline, enhancing efficiency while preserving artistic intent. Additionally, immersive editing tools~\cite{zhou2024timetunnel} tackle VR motion editing by merging spatial and temporal control into a unified interface through keyposes and trajectories, allowing intuitive, direct manipulation of complex 3D motion with reduced editing time. These innovations connect static educational content and interactive simulation by leveraging computer vision (segment anything~\cite{kirillov2023segment}) and multimodal LLMs to extract textbook diagrams and transform them into embedded, interactive physics simulations, enhancing learning engagement~\cite{gunturu2024augmented}.

Sketch animation tools progressively reduce manual workload while maintaining creative control, incrementally automating the animation process, from physical effects and gesture replication to texture generation, semantic interpretation, and procedural motion, moving from assisting traditional animation to creating dynamic, adaptable animations with minimal artist input.
Further advancements were made with progressive layering of control, automation, and interactivity. They evolve from introducing procedural motion into static sketches toward stroke-based specification of spatial-temporal movement and particle-driven dynamic effects, to stylization and exaggeration tools, and real-time performance integration. Later systems expand into specialized workflows like motion loops and novice-friendly static image animation. Collectively, they trace a shift from isolated effect generation to integrated frameworks that combine user-driven artistic input with procedural, physical, and interactive systems, pushing sketch-based animation toward being both expressive and accessible in live, interactive, and production contexts. Further recent methods lie in their shift toward multimodal, immersive, and AI-assisted authoring, expanding from earlier sketch-based animation systems into real-world integration, multimodal storytelling, production-ready AI augmentation, immersive motion control, and domain-specific intelligent content transformation.  


\section{Comparative analysis of animation techniques}
\label{sec:fifth}
We discuss various aspects of sketch animation for each technique and present a comparative study. Specifically, we evaluate suitability across animation types and usage, highlighting when an approach is preferable and where trade-offs emerge. This analysis identifies which methods are effective under challenging conditions and which are sensitive to stroke ambiguity, occlusion, topological changes, or interactions among multiple objects.

\begin{table*}
\centering
\caption{Comparative analysis of sketch animation techniques with attribute-supported citations}
\renewcommand{\arraystretch}{1.45}
\label{tab:analysis}
\scalebox{0.8}{
\begin{tabular}{|p{3.5cm}|p{2.7cm}|p{2.7cm}|p{2.2cm}|p{2.2cm}|p{2.2cm}|p{2.7cm}|}
\hline
\textbf{Attributes / Techniques} 
& \textbf{Inbetweening approach} 
& \textbf{Physics-based} 
& \multicolumn{3}{c|}{\textbf{Data-driven}} 
& \textbf{2D sketch-to-3D model animation} \\ 

\hhline{~~~|---|~} 

 & & & \textbf{MoCap} & \textbf{Video-driven} & \textbf{Text-driven} & \\ 
\hline

\textbf{User Control} 
& High \cite{whited2010betweenit,xing2015autocomplete,brodt2024skeleton}
& Moderate \cite{zhu2011sketch,scott2013physink,guay2015adding}
& Low \cite{smith2023method}
& Low \cite{xie2024video,rai2024sketchanim}
& Low \cite{liu2025dynamic,wu2025aniclipart, deng2025animatesketches, bandyopadhyay2025flipsketch} 
& Moderate \cite{feng2017magictoon, dvorovzvnak2020monster}\\
\hline

\textbf{Automation} 
& Low \cite{whited2010betweenit,narita2019optical}
& High \cite{zhu2011sketch,scott2013physink,li2023enchantedbrush}
& High \cite{dadfar20213a2a,smith2023method}
& High \cite{xie2023wakey,xie2024video}
& High \cite{gal2024breathing,deng2025animatesketches} 
& High \cite{zhou2024drawingspinup, yoon2025occlusion}\\
\hline

\textbf{Generalization} 
& Low \cite{xing2015autocomplete}
& Moderate \cite{lingens2020towards}
& Moderate \cite{patel2016tracemove}
& Moderate \cite{yang2024sketchanimator,rai2024sketchanim}
& High \cite{liu2025dynamic,liang2025multi} 
& Moderate \cite{feng2017magictoon, zhou2024drawingspinup}\\
\hline

\textbf{Temporal Coherence} 
& High \cite{narita2019optical,shen2024bridging}
& High \cite{scott2013physink,eroglu2018fluid}
& Moderate \cite{dvorovznak2018toonsynth}
& Moderate \cite{zhu2025vector}
& Moderate \cite{gal2024breathing, bandyopadhyay2025flipsketch, khandelwal2025flexiclip} 
& Moderate \cite{weng2019photo, yoon2025occlusion}\\
\hline

\textbf{Realism} 
& High \cite{narita2019optical,zhu2025thin}
& High \cite{guay2015adding,suzuki2020realitysketch}
& High \cite{smith2023method}
& High \cite{xie2024video,rai2024sketchanim}
& Moderate \cite{wu2025aniclipart, liang2025multi} 
& Moderate \cite{zhou2025rigging}\\
\hline

\textbf{Complexity} 
& Moderate \cite{xing2015autocomplete, mo2024joint}
& High \cite{li2023enchantedbrush,lingens2020towards}
& Moderate \cite{dadfar20213a2a,smith2023method}
& Moderate \cite{xie2024video}
& Low \cite{liu2025dynamic,gal2024breathing, liu2025multi} 
& Moderate \cite{smith2025animating, dvorovzvnak2020monster}\\
\hline

\textbf{Flexibility} 
& Moderate \cite{brodt2024skeleton}
& Low \cite{zhu2011sketch}
& Moderate \cite{patel2016tracemove}
& Moderate \cite{yang2024sketchanimator}
& High \cite{deng2025animatesketches,liang2025multi} 
& Low \cite{zhou2024drawingspinup}\\
\hline

\textbf{Speed} 
& Moderate \cite{whited2010betweenit}
& Low \cite{scott2013physink}
& Moderate \cite{smith2023method}
& Moderate \cite{su2018live}
& Low \cite{liu2025dynamic, rai2024enhancing, deng2025animatesketches} 
& Low \cite{yoon2025occlusion}\\
\hline

\textbf{Expressiveness} 
& High \cite{shen2024bridging,brodt2024skeleton}
& High \cite{li2023enchantedbrush}
& Moderate \cite{dvorovznak2018toonsynth}
& Moderate \cite{su2018live, rai2024sketchanim}
& Low \cite{wu2025aniclipart, gal2024breathing} 
& Moderate \cite{feng2017magictoon, zhou2025rigging}\\
\hline

\textbf{Abstraction Level} 
& Low \cite{narita2019optical}
& High \cite{zhu2011sketch,eroglu2018fluid}
& Low \cite{smith2023method}
& Moderate \cite{zhu2025vector}
& High \cite{liu2025dynamic,gal2024breathing, bandyopadhyay2025flipsketch} 
& Low \cite{dvorovzvnak2020monster} \\
\hline

\end{tabular}
}
\end{table*}

\paragraph*{User control}
User control refers to the extent of manual intervention and precision that an animator possesses during the animation process. Keyframe-based interpolation~\cite{xing2015autocomplete, whited2010betweenit, brodt2024skeleton} provides strong control, enabling users to define essential poses and adjust timing. In contrast, physics-based techniques~\cite {scott2013physink, guay2015adding, zhu2011sketch} provide moderate control, as motion generation depends on predefined physical constraints and simulation parameters rather than user manipulation. At the same time, data-driven methods, including MoCap~\cite{smith2023method}, video-driven~\cite{rai2024sketchanim, xie2024video}, and text-driven systems~\cite{xie2024video, liu2025dynamic, deng2025animatesketches}, significantly reduce user control, with the animation system making most motion decisions based on input data or natural language descriptions. Similarly, 2D sketch–to–3D model animation techniques~\cite{feng2017magictoon, dvorovzvnak2020monster} provide moderate control, as user input defines sketch drawing and pose while motion generation is partially automated.

\paragraph*{Automation}
This feature captures how much of the animation process is automated by the system. Keyframe-based interpolation methods~\cite{whited2010betweenit, narita2019optical} offer low automation as the system primarily assists with generating inbetween frames based on manually defined key poses. In contrast, physics-based animation techniques~\cite{scott2013physink, guay2015adding, li2023enchantedbrush} achieve high automation by simulating motion dynamics controlled by physical laws, enabling realistic effects such as collisions, elasticity, and inertia with minimal user intervention. Data-driven methods, such as motion-capture-based systems~\cite{smith2023method, dadfar20213a2a} and video-driven methods~\cite{xie2023wakey, xie2024video}, are fully automated, as motion patterns are learned from real-world performance data or directly captured from video, respectively. Similarly, text-driven methods~\cite{gal2024breathing, deng2025animatesketches} are highly automated, relying on generative models to synthesize motion from natural-language descriptions. 2D sketch to 3D models \cite{zhou2024drawingspinup, yoon2025occlusion} where predefined motions often derived from motion-capture or video data are automatically mapped onto 3D models based on user-provided sketches.

\paragraph*{Generalization}
Generalization refers to an animation system's ability to adapt to diverse characters, motion styles, and drawing variations without retraining or supervision from specific animation sources or datasets. Keyframe-based interpolation systems~\cite{xing2015autocomplete} exhibit limited generalization, as motion retargeting is typically required to transfer animations across characters with different proportions or skeletal structures. Physics-based methods~\cite{lingens2020towards} offer moderate generalization, since their adaptability depends on consistent physical parameters and accurate geometric representations, which may not generalize well across stylized or abstract characters.
Data-driven approaches such as MoCap~\cite{patel2016tracemove} and video-driven methods~\cite{yang2024sketchanimator, rai2024sketchanim} depend on how closely the target motion resembles the source during motion transfer, offering moderate generalization, while text-driven methods~\cite{liu2025dynamic, liang2025multi} show strong generalization through generative models that can synthesize motion from abstract text descriptions. However, their performance is bound by the variety and quality of training data. Finally, 2D sketch–to–3D animation techniques~\cite{feng2017magictoon, dvorovzvnak2020monster, zhou2024drawingspinup} offer moderate generalization, as sketch-based inputs can accommodate diverse drawing styles, while the underlying 3D representations still impose structural and rigging constraints.

\paragraph*{Temporal coherence}
Temporal coherence measures the smoothness and consistency of motion over time. Keyframe-based interpolation methods~\cite{shen2024bridging, narita2019optical} achieve high temporal coherence by explicitly enforcing smooth transitions between user-defined key poses. Physics-based method~\cite{scott2013physink, eroglu2018fluid} offers high temporal coherency by simulating motion under continuous physical properties, creating naturally smooth trajectories over time. Data-driven methods such as MoCap~\cite{dvorovznak2018toonsynth} and video-based methods~\cite{zhu2025vector} maintain temporal consistency by following subsequent video pose changes, providing moderate coherency. Text-driven animation systems~\cite{gal2024breathing, bandyopadhyay2025flipsketch, khandelwal2025flexiclip} provide moderate coherence, depending on the underlying model's capacity (text-to-video diffusion) to maintain temporal structure across frames. Finally, 2D sketch–to–3D model animation techniques~\cite{weng2019photo, yoon2025occlusion} offer moderate temporal coherence, as the 3D model animation relies on data-driven motion sources.

\paragraph*{Realism}
Realism evaluates the perceptual quality of animation, reflecting how closely the generated motion resembles real-world behavior. Keyframe-based animation~\cite{shen2024bridging, zhu2025thin} achieves high realism with well-placed keyframes and motion trajectories. Physics-based animation techniques~\cite{guay2015adding, suzuki2020realitysketch} typically produce highly realistic motion by explicitly simulating physical interactions, forces, and material properties, ensuring that movement adheres to real-world dynamics. Data-driven approaches, including MoCap~\cite{smith2023method} and video-driven methods~\cite{rai2024sketchanim, xie2024video}, achieve high realism by often using motion patterns derived from captured or annotated real-world data. Text-driven animation methods~\cite{liang2025multi, wu2025aniclipart} generally offer moderate realism, which depends on the specificity and quality of textual prompts as well as the expressive capacity of the underlying generative models. Finally, 2D sketch–to–3D model animation techniques~\cite{zhou2025rigging} provide moderate realism, as the mapping from abstract sketches to 3D geometry often relies on simplified rigs and predefined motion templates, limiting the fidelity of fine-grained motion details and physical interactions.

\paragraph*{Complexity}
This attribute reflects the technical effort and cognitive load required to create animations. Keyframe-based~\cite{xing2015autocomplete, mo2024joint} methods require key poses as input, resulting in moderate overall effort despite the technical complexity of pose design and timing control. Physics-based methods~\cite{lingens2020towards, li2023enchantedbrush} are more complex to maintain the physical behavior of the sketch motion. Data-driven methods, including MoCap~\cite{smith2023method, dadfar20213a2a}
and video-driven techniques~\cite{xie2024video} exhibit moderate complexity due to the need for extensive data preprocessing, feature extraction, and motion mapping pipelines. Text-driven systems~\cite{gal2024breathing, liu2025dynamic, liu2025multi}, conceptually simple to use, depend on generative models, and generate text-aligned animation without any complexity. Finally, 2D sketch–to–3D model animation techniques~\cite{smith2025animating, dvorovzvnak2020monster} present moderate complexity, as they balance intuitive sketch-based interaction with the additional effort required for 3D reconstruction, rigging, and motion transfer.

\paragraph*{Flexibility}
Flexibility assesses how well an animation method adapts to different use cases, motion styles, or character designs. Keyframe-based interpolation technique~\cite{brodt2024skeleton} offers moderate flexibility, allowing artists to customize timing and poses within a constrained structure. In contrast, Physics-based approaches~\cite{zhu2011sketch} are less flexible, as they are bound by rule-based motion generation and parameter constraints. In comparison, MoCap~\cite{patel2016tracemove} and video-driven methods~\cite{yang2024sketchanimator} demonstrate moderate flexibility, performing well within the scope of captured or learned motions but showing limited adaptability to unseen styles or novel character configurations. Text-driven systems~\cite{deng2025animatesketches, liang2025multi} are highly flexible through prompt-based variety, allowing a wide range of animations to be generated from text descriptions. Finally, 2D sketch–to–3D model animation techniques~\cite{zhou2024drawingspinup} generally offer low flexibility, as they rely on predefined rigging structures and motion mappings that restrict adaptation to significantly different characters or animation styles.

\paragraph*{Speed}
Speed measures the time required by an animator or system to generate a usable animation output. Keyframe-based animation methods~\cite{whited2010betweenit} achieve moderate speed, as interpolation between manually defined poses accelerates motion generation but remains constrained by the amount of user input and editing time. In contrast, physics-based methods~\cite{scott2013physink} are computationally intensive, as they rely on continuous physical simulations and rule-based solvers, leading to slower performance. Data-driven approaches (MoCap and video-driven)~\cite{smith2023method, su2018live} are moderate, as motion data is extracted and transferred to the sketch. At the same time, text-driven animation~\cite{liu2025dynamic, rai2024enhancing, deng2025animatesketches} uses a text-to-video diffusion model, which requires significant training time to estimate motion and align it to animate the sketch, resulting in low speed. Finally, 2D sketch–to–3D model animation techniques~\cite{yoon2025occlusion} also exhibit low speed, due to the additional cost of 3D reconstruction, occlusion handling, rigging, and motion mapping.

\paragraph*{Expressiveness}
Expressiveness captures the dynamic, stylistic richness, or emotional variation that an animation method can convey. Keyframe-based interpolation systems~\cite{shen2024bridging, brodt2024skeleton} enable high expressiveness, as animators can manually compose exaggerated poses, subtle timing, and stylized motion arcs. Physics-based methods~\cite{zhu2011sketch, li2023enchantedbrush} achieve high expressiveness by simulating natural secondary dynamics, such as squash, stretch, and fluid-like deformations, enabling visually rich motion effects. Data-driven methods, such as MoCap~\cite{dvorovznak2018toonsynth}, have moderate expressiveness, as they mainly replicate physical motion without much abstraction, whereas video-driven methods~\cite{su2018live, rai2024sketchanim} offer low expressiveness, as they primarily depend on video motion. In contrast, text-driven approaches~\cite{wu2025aniclipart, gal2024breathing} often exhibit limited expressiveness, as generative models prioritize semantic accuracy and motion plausibility at the expense of stylistic diversity, thereby complicating fine-grained artistic control. Finally, 2D sketch–to–3D model animation techniques~\cite{feng2017magictoon, zhou2025rigging} offer moderate expressiveness, as secondary motions and stylized deformations can be incorporated, but are constrained by predefined rigs, motion templates, and the fidelity of sketch-to-3D reconstruction.

\paragraph*{Abstraction level}
Abstraction refers to the symbolic or stylized characteristic of an animation, reflecting how far the motion departs from real-world behavior. Keyframe-based interpolation methods~\cite{fu2011animated, narita2019optical} exhibit low abstraction, as motion is typically represented through explicit pose-to-pose transitions with limited stylistic deviation. Physics-based methods~\cite{eroglu2018fluid, li2023enchantedbrush} demonstrate higher abstraction, as physical simulations often produce non-literal or exaggerated effects driven by material properties and fluid-like dynamics. Data-driven techniques such as MoCap systems~\cite{dadfar20213a2a} exhibit the least abstraction, replicating real-world motion with minimal interpretive variation, while video-driven methods~\cite{zhu2025vector} provide moderate abstraction through motion stylization and representation preferences during transfer. In contrast, text-driven animation~\cite{liang2025multi, liu2025dynamic, bandyopadhyay2025flipsketch, gal2024breathing} supports high abstraction, capable of generating motion for the different levels of input sketch abstraction based on the input prompt. Finally, 2D sketch–to–3D model animation techniques~\cite{dvorovzvnak2020monster} typically exhibit low abstraction, as they rely on predefined 3D structures and motion mappings that constrain stylization and deviation from representation.

\section{Datasets and evaluation metrics}
\label{sec:sixth}

\subsection{Sketch datasets}
Over the years, several datasets have been introduced to advance research in sketch understanding, each focusing on different aspects of sketch representation as given in Table~\ref{tab:sketch_dataset}. 
TU-Berlin dataset~\cite{eitz2012humans} is widely used, containing 20,000 samples across 250 categories. It captures arbitrary objects through static 2D representations, making it useful for sketch classification and retrieval tasks but lacks stroke sequence information. Another dataset, QuickDraw~\cite{ha2017neural}, provides over 50 million vector sketches with temporal stroke data, offering significant opportunities for large-scale recognition, sequence modeling, and generative tasks. However, the quick-drawing constraint results in noisy and often oversimplified sketches. To address fine-grained understanding, SketchSeg-10K~\cite{wang2019spfusionnet} introduces fine-grained and open-source pixel-wise segmented sketches for part-level recognition, although its category diversity remains limited. It contains 10,000 samples across 10 classes. SketchFix-160~\cite{sarvadevabhatla2017object} pairs noisy sketches with their clean counterparts for sketch correction tasks, but suffers from a small sample size. It contains 3904 sketch samples across 160 categories. The Sketchy dataset~\cite{sangkloy2016sketchy} bridges sketches and photographs, enabling research in cross-modal retrieval like sketch-based image retrieval. Although imperfect alignment between sketches and photos limits its effectiveness, it has 125 categories and 75,471 sketches of 12,500 objects. The SketchyScene dataset~\cite{zou2018sketchyscene} addresses the lack of richly annotated scene-level sketches, particularly for training and evaluating models that understand or generate structured sketches. It contains around 29,056 scene-level sketches, 7000+ pairs of scene templates and photos, and 11000+ object sketches. Animated Drawings~\cite{smith2023method} contains pixel-based hand-drawn sketch characters useful for animation tasks moving toward dynamic sketch representations. It contains over 178,000 amateur drawings of human figures since it is limited to bipedal, and most sketches are noisy.
SketchAnim~\cite{rai2024sketchanim} introduces pixel-based 50 hand-drawn sketches with wide sketch types, including biped, quadruped, and inanimate categories.

Despite these advances, most datasets primarily operate in 2D space, and 3D-aware sketch datasets remain an open research frontier. These datasets outline a clear transition from static to temporal, dynamic sketches, broadening the landscape of sketch-based modeling and generation.

\begin{table*}[htbp]
\centering
\caption{List of sketch datasets.}
\label{tab:sketch_dataset}
\scalebox{1.0}{
\begin{tabular}{l l l l l cc}
\toprule
\textbf{Dataset} & \textbf{Representation} & \textbf{Sketch samples} & \textbf{Category} & \textbf{2D/3D} & \multicolumn{2}{c}{\textbf{Input type}}\\ 
\cmidrule(lr){6-7} \\
 & & & & & \textbf{Pose} & \textbf{Arbitrary} \\ 
\midrule
TU-Berlin~\cite{eitz2012humans} & Stroke & 20K & 250 & 2D Object& & \checkmark \\ 
Quickdraw~\cite{ha2017neural} & Stroke & 50M+ & 345 & 2D Object&  & \checkmark  \\ 
SketchSeg-150K~\cite{qi2019sketchsegnet+} & Stroke & 150K & 20 & 2D Object & \checkmark & \\ 
SketchFix-160~\cite{sarvadevabhatla2017object} & Stroke & 3904 & 160 & 2D Object& \checkmark & \\ 
Sketchy~\cite{sangkloy2016sketchy} & Stroke & 75K & 125 & 2D Object & & \checkmark \\ 
SketchyScene~\cite{zou2018sketchyscene} & Pixel & 29K & -- & 2D Scene & \checkmark & \\ 
Amateur Drawings Dataset~\cite{smith2023method} & Pixel & 177K+ & -- & 2D Object& \checkmark & \\ 
SketchAnim Dataset~\cite{rai2024sketchanim} & Pixel & 50 & -- & 2D Object&  & \checkmark \\ 
\bottomrule
\end{tabular}%
}
\end{table*}

\subsection{Evaluation metrics}
The evaluation of the sketch animation quality captures not only the fidelity of individual frames but also the temporal consistency, smoothness, and perceptual features of the animation sequence. Sketch animation models require assessing the visual quality of generated sequences and the temporal coherence of motion across frames. Overall, the practical evaluation of sketch animation requires a balanced combination of evaluation metrics that offer reproducible quantitative analysis and subjective human evaluations, which capture the perceptual qualities essential for the practical use case. The evaluation matrices for sketch animation are categorized based on image/frame quality, stroke/contour accuracy, temporal consistency, and animation-specific characteristics.

\subsubsection{Image/Frame quality metrics}
These metrics evaluate the quality of individual frames within an animation, often by comparing a generated frame $\hat{I}\in \mathbb{R}^{h\times w}$ to a ground-truth reference image $I\in \mathbb{R}^{h\times w}$.

\paragraph*{Fréchet inception distance (FID)}
FID metric~\cite{heusel2017gans} measures the distribution similarity between real and generated sketch sequences. A lower FID score indicates that the generated sketches are more similar to real sketches in terms of visual quality. It evaluates the effectiveness of the overall realism of the generated frames but does not directly assess temporal smoothness across frames.
\begin{equation}
\mathrm{FID} =
\|\boldsymbol{\mu}_r - \boldsymbol{\mu}_g\|_2^2
+
\mathrm{Tr}
\left(
\boldsymbol{\Sigma}_r
+
\boldsymbol{\Sigma}_g
-
2(\boldsymbol{\Sigma}_r \boldsymbol{\Sigma}_g)^{\frac{1}{2}}
\right),
\end{equation}
where ($\mu_r$, $\Sigma_r$) and ($\mu_g$, $\Sigma_g$) denote the mean vectors and covariance matrices of the real sketch visual feature vectors and the generated sketch feature vectors, $Tr$ denotes the trace of the metrics, respectively.

\paragraph*{Peak signal-to-noise ratio (PSNR)}
PSNR~\cite{gonzalez2008dip} measures the ratio between maximum possible strength of a signal and strength of corrupting noise. In image evaluation, it is used to quantify the difference between two images at the pixel level. A higher PSNR value indicates a more accurate reconstruction, indicating the generated frame is closer to the ground-truth image. 
\begin{equation}
\mathrm{PSNR} =
10 \log_{10}
\left(
\frac{\mathrm{MAX}_I^2}{\mathrm{MSE}}
\right),
\end{equation}
where $\mathrm{MAX}_I$ is the maximum possible pixel value, and MSE is the mean square error between the original and reconstructed signals.

\paragraph*{Structural similarity index (SSIM)}
Unlike PSNR, which only considers pixel-level differences, SSIM~\cite{wang2004ssim} evaluates the structural similarity between two images. It considers illumination, contrast, and structure, making it a better predictor of perceived image quality. A score closer to 1.0 indicates higher similarity.
\begin{equation}
\mathrm{SSIM}(I, \hat{I}) =
\frac{(2\mu_I \mu_{\hat{I}} + C_1)(2\Sigma_{I\hat{I}} + C_2)}
{(\mu_I^2 + \mu_{\hat{I}}^2 + C_1)
(\sigma_I^2 + \sigma_{\hat{I}}^2 + C_2)},
\end{equation}
where $(\mu_{I}, \mu_{\hat{I}})$, $(\sigma_{I^{2}}, \sigma_{\hat{I}^{2}})$, $(\Sigma_{I, \hat{I}})$, and $(C_{1}, C_{2})$ denote the mean intensities, variances, covariance, and stability constants of the ground truth and the generated sketch, respectively.

\paragraph*{Learned perceptual image patch similarity (LPIPS)}
LPIPS~\cite{zhang2018lpips} measures the distance between feature representations of two images extracted from a pre-trained neural network (like VGG~\cite{simonyan2015vgg} or AlexNet~\cite{krizhevsky2012imagenet}). By comparing images in perceptual space, LPIPS better captures human similarity estimation, as it is less sensitive to minor pixel shifts and more focused on overall visual content.
\begin{equation}
\mathrm{LPIPS}(I, \hat{I}) =
\sum_{l}
\frac{1}{h_l w_l}
\sum_{h,w}
\left\|
\mathbf{W}_l \odot
\left(
\tilde{\boldsymbol{\phi}}_l(I)_{h,w}
-
\tilde{\boldsymbol{\phi}}_l(\hat{I})_{h,w}
\right)
\right\|_2^2 ,
\end{equation}

where $\tilde{\phi_l}(\cdot)$ denotes the channel-wise $\ell_2$-normalized features at each spatial location, 
$h_l$ and $w_l$ are the height and width of the feature map at layer $l$, 
$h$ and $w$ index spatial positions, 
$\mathbf{W}_l$ are learned per-channel weights, 
and $\odot$ denotes element-wise multiplication.

\subsubsection{Stroke/Contour accuracy}
These metrics are specific to the unique characteristics of sketches, focusing on the accuracy of the lines and contours. Let the ground-truth stroke set be denoted by $S = {(s_i)}_{i=1}^N, s\in \mathbb{R}^2$ and the generated stroke set by
$\hat{S} = {(\hat{s}_j)}_{j=1}^M, \hat{s}\in \mathbb{R}^2$.

\paragraph*{Chamfer distance (CD)}
Chamfer Distance~\cite{barrow1977chamfer} estimates the average closest distance between two sets of points or curves. In sketch animation, it is used to evaluate how closely a set of generated strokes matches the ground-truth stroke set. A lower CD value indicates a better match.
\begin{equation}
\mathrm{CD}(S, \hat{S}) =
\frac{1}{|S|}
\sum_{s \in S}
\min_{\hat{s} \in \hat{S}}
\| s - \hat{s} \|_2
+
\frac{1}{|\hat{S}|}
\sum_{\hat{s} \in \hat{S}}
\min_{s \in S}
\| \hat{s} - s \|_2 .
\end{equation}

\paragraph*{Hausdorff distance (HD)} 
Hausdorff Distance~\cite{huttenlocher1993hausdorff} estimates the maximum distance of a point in one set to the nearest point in the other set. It is sensitive to outliers or large deviations in the generated strokes, making it a robust metric for ensuring high accuracy and no significant errors.
\begin{equation}
\mathrm{HD}(S, \hat{S}) =
\max
\left\{ 
\max_{s \in S}
\min_{\hat{s} \in \hat{S}}
\| s - \hat{s} \|_2,
\; 
\max_{\hat{s} \in \hat{S}}
\min_{s \in S}
\| \hat{s} - s \|_2
\right\} .
\end{equation}

\subsubsection{Temporal consistency}
Temporal coherence evaluates the smoothness and consistency of the animation sequence. It assures the drastic pose change between consecutive frames that helps to avoid jittery or abrupt transitions. Temporal consistency metrics are essential for evaluating smooth, non-jitter, and realistic animation flows from one frame to the next. 

\paragraph*{Temporal warping error}
Temporal warping measures how well a frame $\hat{I}_{t+1}$ aligns with the previous frame $\hat{I}_t$ when warped using the estimated optical flow~\cite{liu2017videoframe}. A lower warping error implies smoother transitions and reduced temporal jitter.
\begin{equation}
\mathcal{L}_{\mathrm{warp}} =
\frac{1}{T-1}
\sum_{t=1}^{T-1}
\frac{1}{h \cdot w}
\left\|
\hat{I}_t
-
W(\hat{I}_{t+1}, \hat{\mathbf{F}}_t)
\right\|_2^2 ,
\end{equation}
where $\mathcal{L}_{\mathrm{warp}}$ is the warping error, $\hat{F}_t$ denotes optical flow from $\hat{I}_{t}$ to $\hat{I}_{t+1}$, $T$ denotes the total number of frames in the video sequence and $W(\hat{I}_{t+1}, \hat{\mathbf{F}}_t)$ represents backward warping of frame $\hat{I}_{t+1}$ using the estimated flow.

\paragraph*{Fréchet video distance (FVD)}
Similar to FID, FVD~\cite{unterthiner2019fvd} is a metric for evaluating the realism of video sequences. It measures the distance between the feature distributions of real and generated videos. A low FVD score indicates that the generated video sequence is visually and temporally realistic, capturing the actual dynamics of real motion.

\subsubsection{Animation-specific quality metrics}
\paragraph*{Sketch-to-video consistency}
Sketch-to-video consistency~\cite{gal2024breathing} is primarily use for text-driven sketch animation. It estimates the consistency of the generated sketch animation with the input sketch. It uses the CLIP~\cite{radford2021learning} model and estimates the cosine similarity between the input and generated animated sketches. Higher values indicate stronger visual consistency between the input sketch and the generated animation.
\begin{equation}
\mathcal{S}_{\mathrm{SV}} =
\frac{1}{T}
\sum_{t=1}^{T}
\frac{
\langle
\phi_{\mathrm{img}}(I_{\mathrm{ref}}),
\phi_{\mathrm{img}}(\hat{I}_t)
\rangle
}{
\|
\phi_{\mathrm{img}}(I_{\mathrm{ref}})
\|_2
\|
\phi_{\mathrm{img}}(\hat{I}_t)
\|_2
}
\end{equation}
where $\mathcal{S}_{\mathrm{SV}}$ is sketch-to-video consistency score and $\langle$ $\cdot$ $\rangle$ denotes the inner product (dot product). $I_\mathrm{ref}$ denotes the input reference sketch, $\hat{I}_t$ denotes the generated sketch animation at time $t$, $T$ denotes the total number of frames in the generated sketch animation, and $\phi_{img}(\cdot)$ denotes the CLIP image encoder.

\paragraph*{Text-to-video alignment}
Text-to-video alignment~\cite{gal2024breathing} estimates the alignment of the generated sketch animation to the input text prompt $\mathcal{T}_{text}$. It uses the X-CLIP~\cite{ni2022expanding} model, trained on video recognition, to estimate the alignment score. Higher alignment scores indicate better semantic correspondence between the text prompt and the generated sketch animation.
\begin{equation}
\mathcal{S}_{\mathrm{TV}} =
\frac{
\left\langle
\phi_{\mathrm{text}}(\mathcal{T}_{\mathrm{text}}),
\phi_{\mathrm{vid}}(\{\hat{I}_t\}_{t=1}^{T})
\right\rangle
}{
\|
\phi_{\mathrm{text}}(\mathcal{T}_{\mathrm{text}})
\|_2
\|
\phi_{\mathrm{vid}}(\{\hat{I}_t\}_{t=1}^{T})
\|_2
}
\end{equation}
where $\mathcal{S}_{\mathrm{TV}}$ is text-to-video alignment score, $\phi_{text}$ denotes the input text prompt, $\phi_{vid}(\cdot)$ and and $\phi_{text}(\cdot)$ denotes the X-CLIP video and text encoder, respectively.

\subsubsection{User/Subjective evaluation} 
User evaluations are essential for assessing sketch animations because their quality is determined by human perception. It is especially true for sketches, where the goal is often to capture a specific sense or style rather than just technical perfection. In user evaluation, participants are asked to provide feedback for the generated animations based on quality, realism, and consistency. It provides a perceptual validation that automatic metrics may fail to capture fully, especially for subtle attributes like style, expression, and creative interpretation.
Since sketches are highly perceptual, user studies are often required:
\paragraph*{Mean opinion score}
Mean Opinion Score (MOS) is a widely used method for getting human ratings on different aspects of an animation. Participants are typically shown a series of animations and asked to rate them on a scale, often from 1 (poor) to 5 (excellent), for characteristics such as smoothness, quality, and plausibility of motion. The scores from all participants are averaged to get a single, quantifiable MOS for each characteristic, providing insights into how well an animation performs from the user's perspective.
\paragraph*{Pairwise preference tests}
Pairwise preference tests are another powerful way to evaluate animations, especially when comparing two different methods or techniques. Instead of rating animations individually, participants are shown two animations side-by-side and asked to choose which one they prefer, such as, "which animation of a running character looks more natural?" This method is particularly effective because it simplifies the task for the participant. They only have to choose between two options rather than assign a numerical score. Calculating the choices across all participants can determine which method is consistently preferred and by how much, providing a robust comparison between techniques.
\paragraph*{Artist evaluation}
Artist evaluation is crucial for assessing how well an animation tool or technique meets the needs of its intended users. Professional animators, who understand the properties of motion and character performance, are asked to evaluate the system on a variety of factors, including:
\begin{itemize}
    \item \textit{Usability:} How easy and intuitive the tool is to use. Can an artist quickly create what they envision?
    \item \textit{Visual Appeal:} Does the tool allow for creating aesthetically pleasing and expressive animations?
    \item \textit{Workflow Integration:} Can the technique be easily integrated into a professional animation pipeline?
\end{itemize}
This type of feedback is invaluable because it comes from experts who can identify subtle issues. Their insights can directly inform the development of tools that are not only technically sound but also practical and inspiring for the creative professionals who will use them.

Overall, the practical evaluation of sketch animation requires a balanced combination of evaluation metrics that offer reproducible quantitative analysis and subjective human evaluations, which capture the perceptual qualities essential for the practical use case.

\section{Recent trends}
\label{sec:trends}
\paragraph*{Generative animation}
Recent animation methods have seen a significant rise in the use of generative models, particularly diffusion models~\cite{ho2020denoising, song2020denoising, blattmann2023stable}. Previous works relied on Variational Autoencoders (VAEs)~\cite{kingma2013auto} and Generative Adversarial Networks (GANs)~\cite{goodfellow2020generative}. The diffusion model provides an edge by facilitating superior image fidelity and stability during training. It is used for sketch synthesis and animation, offering a more flexible and robust way to generate high-quality outputs from sparse or ambiguous sketch sequences.

Cross-model conditioning has recently been highly explored; generative models use text, video, or pose with the sketch to generate plausible animations. Language-guided sketch animation~\cite{gal2024breathing, liu2025dynamic} enables users to generate sketch animations from text prompts. On the other hand, video-driven models~\cite{hu2024animate, yang2024sketchanimator} animate characters based on keypoint motion or optical flow.
Using foundation models like CLIP~\cite{radford2021learning} and large-scale text-to-image transformers has made such multimodal control more accessible and effective.

\paragraph*{Temporal consistency}
Another emerging trend is using AI models to generate temporally consistent animations. Deep learning-based architectures use temporal conditioning~\cite{ni2023conditional, hu2025animate, rai2024enhancing} to ensure smooth and continuous motion for visual coherency. The temporal coherency in these AI models is conditioned on sketch keyframes or motion cues.

\paragraph*{3D awareness and depth estimation}
3D awareness and depth integration are used to make sketch-based animations more immersive. 2D sketches with depth cues or 3D prior (camera parameters) integration~\cite{rai2024sketchanim} generate an animation that is better suited for augmented reality (AR), virtual reality (VR), and metaverse applications~\cite{feng2017magictoon, smith2025animating}. These models help bridge the gap between artistic 2D inputs and real-world spatial dynamics.

\paragraph*{Learning from sparse data}
In recent years, researchers have increasingly focused on learning from sparse datasets~\cite{maheshwari2023transfer4d}, leading to the adoption of few-shot learning techniques to extend sketch animation models to novel characters or artistic styles with limited training examples. These techniques~\cite{shen2024bridging} are especially valuable given the independency of large-scale, high-quality sketch animation datasets.

\section{Applications}
\label{sec:application}
Sketch animation is an adaptable technology with various applications across diverse domains. Its ability to synthesize dynamic visual narratives from abstract and minimal inputs makes it valuable in creative industries and assistive technologies.

\paragraph*{Creative tools for artists}
Sketch animation technologies empower artists by offering intuitive tools for generating animation content from static illustrations. Users can create dynamic transitions, motion sequences, and visual styles with minimal effort. Creative tools leveraging sketch animation allow for rapid visualization of ideas, iterative storytelling, and stylistic innovation, allowing animation production for novice users and animators.

\paragraph*{Animation in Education and Storytelling}
Sketch animation delivers an intuitive and attractive medium for education and storytelling. In educational settings, sketch animation can facilitate the explanation of complex illustrations by visualizing them dynamically.
Animators or users can use sketch animation techniques to create interactive enhancements with minimalistic and artistic aesthetics. Moreover, sketch-based storytelling is particularly effective for children's educational content, e-books, and scientific illustration animation.

\paragraph*{Medical and scientific visualization}
Sketch animation helps to visualize biological processes, surgical procedures, or physical phenomena in a simplified and dynamic way. Immersive technologies such as AR/VR will be beneficial for visualizing biological processes. It can serve as an educational aid for students, a training resource, or accessible user information.

\paragraph*{Game and character design}
Sketch animation is essential for rapidly prototyping characters, actions, and environments in the game industry. Through animated sketches, artists can quickly iterate on character motion, pose transitions, motion mapping, character stylization, or scene layouts before committing to fully rendered assets. It also enables the dynamic generation of character actions from raw inputs and supports creative exploration and stylistic diversity in game development.

\paragraph*{Virtual avatars and social communication}
Sketch animation helps to create expressive virtual avatars for social media platforms, virtual meetings, and gaming environments. Sketch-based avatars, driven by user inputs or facial expressions, offer a stylized choice to photorealistic avatars, reducing computational costs while maintaining user privacy and personalization. Also, it represents emotions and gestures in a visually appealing, smooth, customizable form.

\paragraph*{Data visualization and infographics}
Animated sketches can be used to bring data visualizations to real-life use cases. Sketch-based animations can dynamically draw out data trends, making information more engaging and easier to understand, rather than creating static charts or graphs. Designers can illustrate building transformations, mechanical operations, or assembly processes in industrial design workflows through animated sketches.

In summary, sketch animation provides a lightweight, expressive, and highly flexible visual language framework that expands its role in education, design, entertainment, storytelling, healthcare, and the arts, highlighting its transformative potential for the future of visual media.

\section{Current limitations and future directions}
\label{sec:future}
Sketch animation poses challenges because sketches are sparse, abstract, and highly variable in style. Despite the advancement in sketch animation using optimization and learning based techniques, creating good quality sketch animation remains a challenging problem. Handling Topology changes, self-occlusions, and partial drawings is challenging, and maintaining temporal coherence while preserving the artist’s objective and sketch quality (thickness and jitter) is nontrivial, especially under non-rigid deformations and articulated motion. Following are prominent challenges in sketch animation.

\paragraph*{Motion coherence}
Temporal coherence across frames poses a significant challenge, which results in inconsistent or jittered motion between frames, thereby reducing the consistency of animations. Recent diffusion-based video models~\cite{chen2023videocrafter1, wang2023modelscope, karras2023dreampose} attempt to enforce temporal consistency but often struggle with sketch data due to its abstract and sparse nature. Generating animations where motion is smooth and objects possess identity across frames continues to be difficult, especially in large motion sequences.

\paragraph*{Lack of large-scale datasets}
 Despite the advancements, sketch animation faces several challenges due to a lack of large-scale and diverse datasets specific to animation tasks. Most methods struggle to generalize across varied sketching styles, often failing when presented with sketches that differ in abstraction, stroke quality, or artistic interpretation from the training data. Generalization and robustness remain open challenges as models fail under slight input perturbations, distinctive poses, or out-of-distribution scenarios. 
 Additionally, most methods are limited to facial expressions or human pose animation. Further, the field suffers from the absence of large-scale, diverse sketch animation datasets, which limits the ability of models to learn generalized representations across different categories, poses, and motion patterns. QuickDraw~\cite{ha2017neural}, TU-Berlin~\cite{eitz2012humans}, or Sketchy~\cite{sangkloy2016sketchy} focus on static sketches and lack temporal annotations or paired modalities such as text, pose, or depth. This scarcity of comprehensive training data limits the scalability and generalization of current models.
 
\paragraph*{Real-time and interactive generation}
In recent times, AI emerged with generative models such as diffusion models~\cite{song2020denoising, song2020score, nichol2021improved, poole2022dreamfusion, blattmann2023stable}, which, although powerful, are computationally intensive and slow during inference, making them unsuitable for interactive generation capabilities and real-time applications such as live animation, educational tools, or games. Methods like distillation or score-based model approximation are still in their early stages for sketch animation.

\paragraph*{Evaluation metrics and benchmarks}
There is a lack of standard evaluation metrics for sketch animation. Current approaches often rely on subjective visual assessments or use metrics from generic video generation tasks, which may not adequately capture the nature of sketch-based content. Establishing benchmark datasets and perceptual metrics specific to sketch animation is essential for driving progress and comparing models relatively.

\paragraph*{Functional relationship in multi-sketch animation}
Multi-sketch animation or sketch storytelling remains a particularly challenging problem in computer graphics and vision. most existing techniques focus on animating single objects, lacking the ability to model interactions and functional relationships between multiple objects, an essential aspect for producing coherent multi-object animations. Earlier, Kitty~\cite{kazi2014kitty} introduced functional relationships for dynamic illustration, but these methods heavily relied on manual user input to explicitly define constraints, which limited scalability and usability. In the progression of AI, multi-sketch animation~\cite{liu2025multi} has emerged, aiming to generate coherent motion and interactions across multiple sketch objects automatically. However, these methods often lack an explicit formulation of functional relationships and struggle to blend automatic spatial reasoning. It fails to generate the multi-object function relationship, leaving a significant gap between manual constraint-driven systems and fully automated, semantically aware sketch animation frameworks.

\paragraph*{Fine-grained control and editing}
Fine-grained control and user editing methods have limited capabilities. Most methods offer limited control over region-specific edits, stroke-level modifications, or animation rapidness. Users have limited control over influencing attributes like style, expression, or motion curves at a granular level. Developing disentangled representations and intuitive user interfaces remains an open research direction.

\subsection{Future directions}
Several promising directions have been pursued for advancement in sketch animation field. One key aspect is developing a real-time sketch animation framework that enables interactive applications to allow artists and animators to receive immediate feedback and adjust dynamically. Similarly, providing fine-grained user control through stroke, motion editing, and semantics can guide motion timing or style, enabling personalized and expressive animations. Another exciting direction can be the multimodal fusion of sketch and audio inputs, which could provide more useful context and semantics for animation, such as animating a sketch based on an audio description of motion. Further enhancement can be the integration of physics-based reasoning, which can significantly enhance physics-based animation, allowing models to maintain object constraints. Multi-object animation is an open challenge, which includes animating multiple objects simultaneously and their functional relationships. An example might be a character cycling, requiring models to learn spatio-temporal coordination and scene dynamics at a higher level. With this advancement, we can create more intelligent, flexible, robust, human-centric sketch animation systems.

\section{Conclusion}
\label{sec:conclusion}
Sketch animation lies at the intersection of artistic expression and computational intelligence, offering a valuable aspect for transforming static hand-drawn sketches into dynamic visuals. In this survey, we have explored the evolution of sketch animation methods, covering traditional techniques, optimization-based methods, deep learning-based approaches, and recent advancements with generative AI. Despite the recent advancements in sketch animation, some challenges remain, including temporal consistency, handling diverse sketching styles, supporting real-time inference, and generalizing across varied scenes and inputs. Moreover, multi-object interactions further constrain current systems. Further,  the development of interactive, multimodal, and context-aware animation pipelines that combine user feedback, physics reasoning, and scene understanding can be emphasized in future research. With these developments, the current challenge can be addressed since sketch animation can evolve into a powerful tool for creators, educators, and storytellers across domains. This paper serves as a comprehensive reference for researchers in sketch animation and a foundation for future research.

\bibliography{main.bib}
\bibliographystyle{ieeetr}

\end{document}